\documentclass[a4paper,fleqn]{cas-sc}

\usepackage[authoryear]{natbib}

\usepackage{threeparttable}
\usepackage{array}
\usepackage{xurl}
\usepackage{longtable}
\usepackage{placeins}
\graphicspath{{./}}

\makeatletter
\def\@fpsadddefault{\edef\@fps{htbp}}
\makeatother

\ExplSyntaxOn
\cs_set:Npn \__make_fig_caption:nn #1#2
{
  \l_fig_align_tl
  \skip_vertical:N \l_fig_abovecap_skip
  \setbox\cascaptionbox=\hbox{%
     \rmfamily\small\textbf{\color{scolor}#1:}~#2}
  \ifdim\the\wd\cascaptionbox<\dim_use:N \l_fig_width_dim\relax
    \parbox{ \l_fig_width_dim }
      {\unskip\ignorespaces\hfil\rmfamily\small
       \textbf{\color{scolor}#1:}~#2\hfil\par }
  \else
    \parbox{ \l_fig_width_dim }
      {\rightskip=0pt\unskip\ignorespaces\rmfamily
       \small\textbf{\color{scolor}#1:}~#2\par }
  \fi
  \skip_vertical:N \l_fig_belowcap_skip
}
\cs_set:Npn \__make_tbl_caption:nn #1#2
{
  \l_tbl_align_tl
  \skip_vertical:N \l_tbl_abovecap_skip
  {\parbox{ \dimexpr(\l_tbl_width_dim)}
    {\rightskip=0pt\rmfamily\small\textbf{\color{scolor}#1}\par#2\par\vskip4pt }}
  \skip_vertical:N \l_tbl_belowcap_skip
}
\ExplSyntaxOff
\makeatletter
\renewcommand{\LT@makecaption}[3]{%
  \LT@mcol\LT@cols {@{}l@{}}{\parbox[t]{\LTcapwidth}{%
    \rightskip=0pt\rmfamily\small\textbf{\color{scolor}#2}\par #3\par\vskip2pt}}}
\makeatother

\newcommand{\nl}{\tabularnewline}
\newcommand{\hdr}[1]{\begin{tabular}[b]{@{}c@{}}#1\end{tabular}}

\newcommand{\limitationsparagraph}{}

\ExplSyntaxOn
\RenewDocumentEnvironment { PrelimsAbstract } { O{} }
  {\parindent=0pt
   { \fontsize{14pt}{16pt}\selectfont #1 }\par
   \vskip12pt
   { \fontsize{12pt}{14pt}\bfseries\selectfont\casprelimstitle } \par
   \vskip6pt
   \ifnum\theblind>0\relax
     \vspace*{\the\baselineskip}
   \else
     \seq_use:Nn \g_stm_prelimsau_seq { ,~ }
   \fi
   \vskip12pt
   \par
  }
  {}
\ExplSyntaxOff

\ExplSyntaxOn
\RenewDocumentCommand \printorcid { }
{
  \seq_if_empty:NF \g_stm_orcid_seq
  {
    \group_begin:
      \tex_let:D \thefootnote \relax \footnotetext
      {
        \raggedright
        \textsc{orcid}(s):\c_space_token
        \seq_use:Nn \g_stm_orcid_seq { ;~ }
      }
    \group_end:
  }
}
\ExplSyntaxOff

\newcommand{\nNarratives}{5,018,079}
\newcommand{\nHitsExpanded}{6,840}
\newcommand{\nHitsNarrow}{5,401}
\newcommand{\nHitsCore}{5,416}
\newcommand{\nHitsNoisyOnly}{1,424}
\newcommand{\prefilterPrevalence}{0.136}
\newcommand{\nConfirmed}{5,333}
\newcommand{\nProbSum}{5,277}
\newcommand{\nAdjustedTotal}{5,467}
\newcommand{\nAdjustedLo}{5,306}
\newcommand{\nAdjustedHi}{5,577}
\newcommand{\nFetalHarm}{58}
\newcommand{\pctFetalHarm}{1.1}
\newcommand{\coreConfirmRate}{97.9}
\newcommand{\noisyConfirmRate}{2.1}
\newcommand{\nRoleDriver}{3,614}
\newcommand{\pctRoleDriver}{67.8}

\newcommand{\pctRolePassenger}{31.5}

\newcommand{\pctRolePedOther}{0.7}
\newcommand{\nStageEarly}{271}
\newcommand{\pctStageEarly}{5.1}

\newcommand{\pctStageMid}{12.5}
\newcommand{\nStageLate}{949}
\newcommand{\pctStageLate}{17.8}

\newcommand{\pctStageNotStated}{64.6}

\newcommand{\pctOutNoComplaint}{16.5}

\newcommand{\pctOutPain}{25.0}

\newcommand{\pctOutTransported}{57.4}

\newcommand{\seVal}{0.999}
\newcommand{\seLo}{0.997}
\newcommand{\seHi}{1.000}
\newcommand{\spVal}{1.000}
\newcommand{\spLo}{1.000}
\newcommand{\spHi}{1.000}
\newcommand{\ppvVal}{1.000}
\newcommand{\prevInHits}{0.782}
\newcommand{\nValStageA}{174}
\newcommand{\nValStageC}{218}
\newcommand{\nStageCConfirmed}{14}
\newcommand{\nStageCReviewed}{42}
\newcommand{\nStageCScreened}{499,306}
\newcommand{\nStageCPositives}{45}
\newcommand{\stageCRate}{9.0}
\newcommand{\nEstMisses}{148}
\newcommand{\nEstMissesPerYear}{142}

\newcommand{\nPosPerYear}{5.0}
\newcommand{\prefilterRecall}{97.3}
\newcommand{\guardPairs}{590}
\newcommand{\guardAgreement}{100.0}
\newcommand{\guardR}{0.998}
\newcommand{\guardMAD}{0.0034}
\newcommand{\nRoleRead}{300}
\newcommand{\nRoleSettled}{284}
\newcommand{\nRoleConfirmed}{284}
\newcommand{\nFetalRead}{58}
\newcommand{\nFetalSettled}{56}
\newcommand{\nFetalConfirmed}{56}

\newcommand{\nStageCReadAll}{1,446}
\newcommand{\nStageCFoundAll}{0}
\newcommand{\nStageCCensus}{546}
\newcommand{\stageCCensusLo}{0.10}
\newcommand{\nStageCTestedPool}{1,883}
\newcommand{\nStageCTestedRead}{846}
\newcommand{\stageCUpperStratum}{13}
\newcommand{\stageCUpperState}{133}
\newcommand{\nStageCUnsettled}{1}

\newcommand{\nFloorBand}{456,920}
\newcommand{\nSecondOrderTerms}{27}
\newcommand{\nFloorFrame}{2}
\newcommand{\nFloorPregnant}{0}

\newcommand{\nRolePresenceSettled}{299}
\newcommand{\nRolePresenceConfirmed}{299}
\newcommand{\rolePPV}{100.0}

\newcommand{\roleAgree}{96.7}
\newcommand{\fetalAgree}{96.6}
\newcommand{\ageStdLo}{1.066}
\newcommand{\ageStdHi}{1.080}
\newcommand{\ageStdYears}{4}
\newcommand{\ageStdGfrGap}{0.19}
\newcommand{\nRowsRead}{5,109,746}
\newcommand{\nExcludedEmpty}{91,175}
\newcommand{\nExcludedShort}{492}
\newcommand{\nShortHits}{0}
\newcommand{\minChars}{40}
\newcommand{\screenChi}{6.52}
\newcommand{\screenChiDf}{8}
\newcommand{\screenChiP}{0.59}
\newcommand{\nUncertainLabels}{15}
\newcommand{\seWorst}{0.9973}
\newcommand{\spWorst}{0.9714}
\newcommand{\missTerm}{148}
\newcommand{\missAtHalf}{296}
\newcommand{\totalAtHalf}{5,615}
\newcommand{\totalComplete}{4,935}
\newcommand{\rateMinComplete}{1.15}
\newcommand{\rateMaxComplete}{1.45}
\newcommand{\survMinComplete}{2.56}
\newcommand{\survMaxComplete}{3.09}
\newcommand{\coverageLastYear}{97.5}
\newcommand{\pctFemaleDrivers}{98.7}
\newcommand{\nFemaleDrivers}{2,857,416}
\newcommand{\nEligibleSole}{1,955,013}
\newcommand{\nDocCases}{2,465}
\newcommand{\nAmbiguousDropped}{1,054}
\newcommand{\nAmbiguousPersons}{2,194}

\newcommand{\nDriverUnmatched}{95}
\newcommand{\rateMin}{1.13}
\newcommand{\rateMax}{1.45}
\newcommand{\survMin}{2.56}
\newcommand{\survMax}{3.09}
\newcommand{\prevPregLo}{4.6}
\newcommand{\prevPregHi}{5.0}
\newcommand{\expectedTotal}{129,297}
\newcommand{\nAdjustedDrivers}{3,596}
\newcommand{\survShortfall}{36}
\newcommand{\expectedMin}{11,774}
\newcommand{\expectedMax}{15,918}
\newcommand{\bootB}{2,000}
\newcommand{\nTermPregnan}{5,318}
\newcommand{\nTermSolePregnan}{5,112}
\newcommand{\nTermExpecting}{864}

\newcommand{\nTermWithChild}{544}

\newcommand{\bandLo}{0.30}
\newcommand{\bandHi}{0.95}
\newcommand{\yearCountMax}{2019}
\newcommand{\yearCountMin}{2020}
\newcommand{\yearRateMax}{2019}
\newcommand{\yearRateMin}{2025}
\newcommand{\yearSurvMax}{2019}
\newcommand{\yearSurvMin}{2023}
\newcommand{\nCountMax}{708}
\newcommand{\nCountMin}{495}
\newcommand{\survLast}{2.57}
\newcommand{\yearLast}{2025}
\newcommand{\ageDoc}{27.0}
\newcommand{\ageAll}{30.8}
\newcommand{\pctInjuredDoc}{73.2}
\newcommand{\pctInjuredAll}{19.5}
\newcommand{\pctAirbagDoc}{26.5}
\newcommand{\pctAirbagAll}{21.5}
\newcommand{\pctEjectedDoc}{0.6}
\newcommand{\pctEjectedAll}{0.4}
\newcommand{\pctUnbeltedDoc}{0.9}
\newcommand{\pctUnbeltedAll}{0.9}
\newcommand{\nTranspStandaloneOnly}{192}
\newcommand{\nTranspGatedOnly}{105}
\newcommand{\nTransportedGatedEarly}{142}
\newcommand{\nTransportedGatedLate}{591}
\newcommand{\cellTP}{5,305}
\newcommand{\cellFN}{5}
\newcommand{\cellFP}{0}
\newcommand{\cellTN}{1,477}
\newcommand{\rrInjured}{3.75}
\newcommand{\rrAirbag}{1.23}
\newcommand{\rrUnbelted}{1.06}
\newcommand{\rrEjected}{1.57}
\newcommand{\pctLateDriver}{18}
\newcommand{\pctNotStatedDriver}{64}
\newcommand{\pctLatePassenger}{18}
\newcommand{\pctNotStatedPassenger}{65}
\newcommand{\cfgPctOne}{16.7}

\newcommand{\cfgPctTwo}{16.5}

\newcommand{\cfgPctThree}{12.6}

\newcommand{\pctPainEarly}{30}

\newcommand{\pctPainMid}{25}

\newcommand{\pctPainLate}{21}
\newcommand{\pctNoComplaintLate}{15}
\newcommand{\transpEarly}{56.8}
\newcommand{\transpMid}{60.5}
\newcommand{\transpLate}{63.9}
\newcommand{\transpAll}{59.0}
\newcommand{\docProbNoInjury}{0.041}
\newcommand{\docProbSerious}{0.022}
\newcommand{\docAUC}{0.814}
\newcommand{\docN}{252,465}
\newcommand{\docClusters}{254}
\newcommand{\orInjured}{21.7}
\newcommand{\orInjuredLo}{15.9}
\newcommand{\orInjuredHi}{29.6}
\newcommand{\orAirbag}{0.440}
\newcommand{\orUnbelted}{0.579}
\newcommand{\orInjuredAny}{0.685}
\newcommand{\orSerious}{0.776}

\newcommand{\orAge}{0.648}
\newcommand{\orAgeLo}{0.625}
\newcommand{\orAgeHi}{0.671}
\newcommand{\trendRR}{0.974}
\newcommand{\trendPct}{2.6}
\newcommand{\trendRRLo}{0.961}
\newcommand{\trendRRHi}{0.989}
\newcommand{\trendP}{0.00012}
\newcommand{\trendRRComplete}{0.976}
\newcommand{\trendRRCompleteLo}{0.961}
\newcommand{\trendRRCompleteHi}{0.992}
\newcommand{\trendAlpha}{0.00025}
\newcommand{\covidRR}{0.898}
\newcommand{\stateCountyRate}{1.26}
\newcommand{\nCounties}{254}
\newcommand{\sevOR}{4.15}
\newcommand{\sevORBinary}{1.62}
\newcommand{\sevORBinaryUnw}{1.62}
\newcommand{\sevORLo}{3.88}
\newcommand{\sevORHi}{4.44}
\newcommand{\biasGridN}{81}
\newcommand{\biasSeriousLo}{0.03}
\newcommand{\biasSeriousHi}{1.46}
\newcommand{\biasEqualLo}{1.60}
\newcommand{\biasEqualHi}{1.66}
\newcommand{\biasDocMin}{0.02}
\newcommand{\poCutMin}{1.54}
\newcommand{\poCutMax}{8.54}
\newcommand{\poCutN}{3}
\newcommand{\poSpread}{5.6}
\newcommand{\sevBiasLo}{0.030}
\newcommand{\sevBiasHi}{88.119}
\newcommand{\nNarrativesRead}{407}
\newcommand{\nUnclearSetAside}{15}
\newcommand{\kappaVal}{0.944}
\newcommand{\nKappaPairs}{212}
\newcommand{\nDoubleRead}{229}

\newcommand{\nConflicts}{21}
\newcommand{\nSingleCoded}{178}
\newcommand{\countyHiName}{Bexar}
\newcommand{\countyHiRate}{3.45}
\newcommand{\countyLoName}{Hidalgo}
\newcommand{\countyLoRate}{0.42}
\newcommand{\countyTopSpread}{8.2}
\newcommand{\countyNearName}{Harris}
\newcommand{\countyNearRate}{1.32}
\newcommand{\rateUrban}{1.37}
\newcommand{\rateRural}{0.93}
\newcommand{\nPIIReviewScreened}{407}
\newcommand{\nPIIReviewFlagged}{152}
\newcommand{\pctPIIReviewFlagged}{37}
\newcommand{\nPIIScreened}{58}
\newcommand{\nPIIExcluded}{25}
\newcommand{\pctPIIExcluded}{43}

\newcommand{\labelSource}{human adjudication}

\begin{document}
\let\WriteBookmarks\relax

\shorttitle{Documented pregnancy in Texas crash narratives}
\shortauthors{A. Rafe and S. Das}

\title[mode=title]{Counting the Uncounted: Population-Level Surveillance of Documented
Pregnancy and Fetal Harm in Police Crash Narratives with a System One Model (Jev)}

\author[1]{Amir Rafe, Ph.D.}[orcid=0000-0002-4089-2088]
\cormark[1]
\ead{amir.rafe@txstate.edu}
\credit{Conceptualization, Methodology, Software, Formal analysis, Data curation, Validation,
Visualization, Writing - original draft, Writing - review \& editing}

\author[1]{Subasish Das, Ph.D.}[orcid=0000-0002-1671-2753]
\ead{subasish@txstate.edu}
\credit{Conceptualization, Methodology, Supervision, Writing - review \& editing}

\affiliation[1]{organization={Civil Engineering, Texas State University},
  addressline={601 University Drive},
  city={San Marcos},
  postcode={78666 TX},
  country={USA}}

\cortext[cor1]{Corresponding author}

\begin{abstract}
Motor-vehicle crashes are a leading cause of traumatic fetal death in the United States, yet no
crash database records pregnancy. Existing estimates come from national investigation samples
with wide intervals, trauma registries seeing only the injured, or record linkage in a few
states. Police narratives record pregnancy whenever it matters at the scene, and no study has
read them at population scale. We read every police crash narrative in Texas from 2017 through
2025, \nNarratives{} in total. A regular-expression screen flags \nHitsExpanded{} candidates,
which Jev, a calibrated System One decision model, reads under an eight-question schema,
returning probabilities and no text. It confirms \nConfirmed{} as documenting a pregnant person
involved in the crash, with role, stage and post-crash condition. The same model reads
\nStageCScreened{} of the remainder under the presence question, to estimate what the
screen missed. Blind human adjudication of a probability-stratified sample gives weighted
sensitivity \seVal{} and specificity \spVal{}, from which the Rogan-Gladen estimator and a
bootstrap give \nAdjustedTotal{} crashes with a documented pregnancy, with a 95\% interval of
\nAdjustedLo{} to \nAdjustedHi{}, including \nFetalHarm{} crashes documenting fetal harm.
Person-level records give a rate of \rateMin{} to \rateMax{} per 1{,}000 female drivers aged 15
to 49. Measured against a live-birth expectation from vital statistics, narratives document
\survMin{}\% to \survMax{}\% of expected pregnant drivers, and documented-case status is most
strongly associated with the driver's own recorded injury, at an odds ratio of \orInjured{}.
\end{abstract}

\begin{highlights}
\item \nNarratives{} Texas crash narratives screened, two strata read by a calibrated model
\item \nAdjustedTotal{} police-reported crashes with a documented pregnancy, 2017 to 2025
\item \rateMin{} to \rateMax{} per 1,000 female drivers aged 15 to 49, falling \trendPct{}\% a year
\item Narratives document \survMax{}\% of the live-birth expectation at most in any year
\item Documented-case status is most strongly tied to the driver's own injury
\end{highlights}


\maketitle

\section{Introduction}\label{sec:intro}

Motor-vehicle crashes are the leading cause of traumatic fetal death in the United States
\citep{weiss2001} and a major cause of injury-related hospitalization during pregnancy
\citep{schiff2005}. The clinical literature has understood the mechanism for decades, and the
obstetric guidance that follows from it is settled enough to appear in routine prenatal care.
What has never been available is a count from routine state records. No state crash
database in the United States codes pregnancy, and the sources that do carry it are a
weighted national investigation sample, a trauma registry, or a record linkage available in
a few states. A state that wants to know how many of its own police-reported crashes last
year involved a pregnant road user, at what stage, in what role and with what documented
outcome, has nowhere in its own files to look.
That absence is not a matter of academic curiosity alone. The question bears directly on how a
crash report form
is designed, on what an emergency medical technician is told to ask at the scene, and on how a
prenatal provider decides whether restraint counseling is worth the minutes it takes.

The sources that do exist answer narrower questions and cannot be stretched to cover this one.
Weighted national crash-investigation samples estimate fetal deaths with intervals wider than
half their point estimates and are built from a sampling frame of towed passenger-vehicle crashes
that excludes most of what a police department records \citep{viano2023}. Trauma registries
observe only patients injured badly enough to be admitted, which is the wrong end of a
distribution in which most crashes involving a pregnant occupant are minor \citep{chan2025}.
Record linkage between birth certificates and crash reports produces exactly the quantity a
surveillance system needs, and it exists in a handful of states, because linkage is an
administrative achievement rather than a methodological one and most states have not made it
\citep{vladutiu2013aap,hyde2003}. None of these sources can say how many pregnant people are
involved in police-reported crashes in a given state in a given year, and none can say what
fraction of those involvements is recorded anywhere at all.

The information that is missing from the coded fields exists in plain text a few lines away.
Police officers write a narrative for every crash they
attend, and they note pregnancy whenever it changes what happens at the scene, whether to call
for emergency medical services, whether to transport, whether to write a citation. A narrative
that reads like a statement of gestational age followed by a complaint of abdominal pain is a
data point about a pregnant road user that no coded field on the same report captures. Officers
have been writing such sentences for as long as they have been writing narratives. What has been
missing is any way to read five million of them.

This paper reads every one of them for the state of Texas across nine years. We use Jev, a System
One model trained with reinforcement learning for calibrated decisions, which answers an
analyst-defined typed question about a piece of text and returns a probability that is meant to
be taken at face value rather than as a ranking score. The model generates no text, so there is no free-form output to parse and the space of
possible answers is fixed before the narrative is read. A typed answer can still be the
wrong answer, which is why the extractor is measured against human adjudication rather than
trusted. Its calibration is what allows a probability threshold to be turned into a count
with an interval attached, and the whole statewide corpus can be read on ordinary research
infrastructure without a dedicated computing allocation. We treat the model as an instrument and we audit
that instrument inside this paper against human adjudication of a probability-stratified sample
of its own output. That audit reports sensitivity, specificity and inter-rater agreement for the
one variable the analysis depends on, rather than assuming that calibration holds because a
companion paper says so.

Reading the narratives is only the first half of the problem, and on its own it would produce a
number that is easy to misuse. A count of narrative mentions is a count of what officers wrote
down, and the gap between that and what occurred is both large and patterned in ways that the
under-reporting literature has documented for other variables. We therefore build the gap into
the design rather than noting it as a caveat. A regular-expression screen splits the corpus
into a small flagged stratum and a very large remainder, and the model answers eight questions
about every flagged narrative. A random sample of the unflagged remainder is then asked the
presence question, so that what the screen missed can be estimated rather than assumed
away. Human
adjudication of a stratified sample supplies the sensitivity and specificity that correct the
flagged-stratum count for classifier error. Person-level crash records supply denominators, so
the result is a rate rather than a count. Texas vital statistics supply the number of pregnant
drivers that would be expected in crashes if pregnancy were recorded whenever it was present, and
the ratio of the documented count to that expectation is what we call the narrative surveillance
sensitivity. A logistic model of which crashes carry a written-down pregnancy then sits beside that
ratio, and both point the same way.

Three questions organize what follows, and each one depends on the answer to the one before it.
The first asks how many pregnant road users a state's police narratives document over a defined
period. That count has to be corrected both for the error rate of the instrument that read the
narratives and for what the screen in front of that instrument missed. The second asks what the
corrected count is a rate of, which requires a denominator of comparable people rather than a
bare total, and how that rate moves over nine years. The third asks how much of the underlying
phenomenon a narrative-based count sees at all. We answer it by comparing the documented count
against the involvement that vital statistics lead one to expect, and then by modeling which
crashes get a pregnancy written down in the first place. A count without an error rate cannot be
compared with anything, a count without a denominator cannot be compared across years, and a
rate without a documentation model cannot be told apart from a change in reporting practice.

Four contributions follow from the design set out above. The first is a validated, calibrated
narrative
extractor for pregnancy involvement, road-user role, gestational stage and documented outcome,
released as an eight-question schema that any state can run against its own narrative file. The
second is nine years of statewide counts and rates corrected for measured classifier error and
for measured screen recall, with every interval carried through a single bootstrap chain rather
than assembled from separately propagated pieces. The third is a documentation model and the surveillance-sensitivity estimate beside it. Those two together convert a count into a
statement about how much of the phenomenon the reporting system sees at all, and they are the
part of this work most worth replicating elsewhere. The fourth is the descriptive epidemiology
that follows, covering role, stage, restraint and airbag use, crash configuration, timing and
geography, together with a documented fetal-harm series reported in aggregate.

One word in this paper carries more weight than any other and is used consistently throughout.
When we write that a pregnancy is \emph{documented}, we mean that an officer wrote something down
in a narrative, which is a fact about the record and not a measurement of prevalence. Every count below is a documentation count, every rate is a documentation rate, and the
distance between documentation and prevalence is not a nuisance at the edge of this study
but the quantity that Section~\ref{sec:res-doc} sets out to measure. The unit of that count
is one police-reported crash whose narrative records a pregnancy, and not one pregnant
person. A binary presence question cannot separate two pregnant people named in the same
narrative, and no step of the design attaches a person-level record to every corrected case,
so a crash carrying more than one pregnancy contributes one to every total in this paper. A reader who substitutes one for the other will
draw conclusions this paper does not support, and a reader who keeps them apart will find that
the gap between them is the most portable result here.

\section{Background}\label{sec:background}

\subsection{The burden of crash trauma in pregnancy and how it has been measured}\label{sec:bg-burden}
Motor-vehicle crashes are the leading mechanism of traumatic fetal death in the United States,
and the studies that established this used the only population source available for the purpose.
Reviewing fetal death certificates from sixteen states covering 55\% of United States live
births, \citet{weiss2001} identified 240 traumatic fetal deaths over three years, a rate of 3.7
per 100{,}000 live births, of which crashes accounted for 82\%. In that study the rate peaked at
9.3 per 100{,}000 among women aged 15 to 19 \citep{weiss2001}. The unit there is a certified fetal death, which is the
right unit for a mortality rate and the wrong one for counting involvement. A meta-analysis of observational studies reports pooled incidences per 1{,}000 women
involved in a crash of 276.43 for induced labor, 191.90 for preterm delivery and 166.65
for cesarean section \citep{amezcuapriet2020maternal}. Against women not involved it reports raised odds of placental abruption at 1.43 and of
maternal death at 202.27 \citep{amezcuapriet2020maternal}. Clinical reviews give the wider frame in
which those rates sit. Trauma affects between six and seven percent of pregnancies and accounts
for as much as 46\% of maternal deaths \citep{chames2008trauma}, and physical trauma reaches one
pregnancy in twelve \citep{jain2015guidelines}. Across fifty-one studies and 95{,}949 patients, trauma
is the leading non-obstetric cause of maternal death and crashes are its commonest mechanism
\citep{petrone2019traumatic}. The practice guideline of \citet{barraco2010practice} puts trauma
at 7\% of pregnancies with admission required in four per 1{,}000, notes that incidence rises
with gestational age and that just over half occurs in the third trimester, and attributes half
of those traumas to crashes. Blunt abdominal trauma is the leading injury type, with abruption,
preterm labor and uterine rupture the outcomes that follow \citep{greco2019guidance}, and a
systematic review of 225 included studies finds the evidence still too thin to settle the safest
management strategy \citep{mendezfigueroa2013trauma}.

The harm is not confined to severe crashes, which is the single fact that most constrains a
surveillance system keyed to injury. In a Washington State cohort of hospitalized pregnant women,
those recording an Injury Severity Score of zero still had sharply elevated risk of preterm labor
at a relative risk of 7.9 with an interval of 6.4 to 9.8. The same uninjured group had elevated
risk of placental abruption at 6.6 with an interval of 3.9 to 11.2 \citep{schiff2005}. \citet{viano2023} found that in 82.9\% of crashes
producing a fetal death the pregnant occupant was not seriously injured, and that the most common
mechanism was a minor crash. A single documented case shows the same thing at the level of one
patient, since a restrained driver with airbag deployment and no major maternal injury went on to
show fetal subdural hematoma and intraventricular hemorrhage \citep{matsushita2014fetal}.
Consequences also appear a long time after the crash itself. An Ontario cohort of 2.1 million children linked
in-utero exposure to maternal unintentional injury to cerebral palsy at an adjusted hazard ratio
of 1.33 with an interval of 1.18 to 1.50, rising when the injury required hospitalization
\citep{ahmed2023in,redelmeier2016motor}. A Taiwanese birth-notification cohort of 19{,}277
crash-exposed offspring against 76{,}015 comparisons estimates the same kind of association for
neonatal outcomes \citep{chang2023maternal}. For this paper the consequence of that pattern is direct. If the
outcomes that matter arise from crashes in which nobody is badly hurt, then any count assembled
from records that are written because somebody was hurt will miss most of the population at risk.

Restraint use is the principal modifiable factor and has been studied from three directions.
\citet{klinich2008} found fetal outcome most strongly associated with crash severity and maternal
injury, with proper belt use associated with acceptable fetal outcome at an odds ratio of 4.5,
and estimated that roughly half of crash-related fetal losses would be preventable under
universal correct belt use. Biomechanical work supports that finding from the laboratory. The first
pregnant crash test dummy showed substantial energy transmission to the uterine compartment under
the restraint configurations pregnant women commonly adopt \citep{pearlman1996crash}, and belted
pregnant-anthropometry dummies showed reduced abdominal pressure and avoided steering-wheel
contact in frontal and rear impacts \citep{motozawa2010effect}. Later work points to side loading
of the abdomen rather than the belt itself as a dominant injury source
\citep{viano2026source,duma2006analysis}. Work on national crash samples finds belts
protective for pregnant occupants with no offsetting injury penalty
\citep{espelien2025injury,hattori2021}, and a clinical series of 126 pregnant patients associates
restraint non-use with increased fetal death after a crash \citep{luley2013seatbelt}. What the
counseling literature then shows is that this knowledge does not reach the people it concerns.
Across nineteen states, counseling to wear a seat belt during pregnancy ranged from 36.7\% to
56.5\% and belt use among reproductive-aged women from 69.5\% to 91.4\%
\citep{beck2005prevalence}. Fewer than half of pregnant women report receiving restraint
counseling from a prenatal provider \citep{sirin2007seat,taylor2005seatbelt}, and among 1{,}491
Australian pregnant drivers 99.1\% wore a belt while only 41.4\% positioned it correctly
\citep{koppel2025assessing}. In a population-based survey of women with a recent live birth, 2\%
were hurt in a crash during pregnancy, only 57\% of those had been counseled on belt use, and
87\% needed medical care \citep{whitehead2013prenatal}. Driving exposure itself changes over the course of a pregnancy.
A self-matched Ontario cohort of 507{,}262 women found a 42\% relative increase in serious crash
risk during the second trimester against the pre-pregnancy baseline, with an interval of 32 to
53\% and no similar increase for crashes as a passenger or pedestrian \citep{redelmeier2014}.
That gradient is a reason to record gestational stage and road-user role rather than pregnancy
alone, and it is why this paper extracts both.

Table~\ref{tab:lit} positions the present study against this literature along the dimensions
that decide whether a source can support surveillance. Those dimensions are the data source, the
unit of analysis, whether a denominator exists, whether stage and fetal outcome are observed,
whether misclassification is adjusted for, and the population scope. The pattern across those
six dimensions is consistent. Sources
with clinical depth lack denominators, as in the trauma-center series of \citet{chan2025} with
157 patients and the national investigation sample of \citet{hattori2021} with 736 injured women,
and a decade of Level-I trauma admissions matched one to three against non-pregnant women shares
that property \citep{owattanapanich2021motor}. Sources with denominators lack narrative detail,
as in the linked birth and crash records of \citet{vladutiu2013aap,vladutiu2013ajpm}, the Utah
linkage of \citet{hyde2003} with 8{,}938 pregnant women in crashes, and the Washington hospital
linkage of \citet{schiff2005}. The Swedish national register linkage of
\citet{kvarnstrand2008maternal} shares that property and found 1.4 maternal fatalities per
100{,}000 pregnancies from crashes between 1991 and 2001. None covers a state population of police-reported crashes.
Record linkage is the established route to a denominator, and its ceiling is administrative
rather than methodological, since only four United States states had linked vital records to
crash reports at the time of the largest such study \citep{vladutiu2013aap}. Where linkage
exists it is informative, and \citet{vladutiu2013aap} estimated a crash risk of 12.6 per 1{,}000
pregnant women in North Carolina across 878{,}546 pregnancies. The companion analysis found that
a second crash raised stillbirth risk at an adjusted relative risk of 4.82 with an interval of
2.85 to 8.14 \citep{vladutiu2013ajpm}. Linkage quality nonetheless varies by jurisdiction and
by casualty characteristics. A meta-analysis of 54 crash-to-hospital linkage studies found widely
varying match rates and warned that linkage rate alone is an inadequate quality measure, because
age, sex, crash type and road geometry all predict whether a record appears in the linked set
\citep{karimi2024a}. Linked systems such as the crash outcome data evaluation system demonstrate
the value of the approach where it is available
\citep{sauberschatz2015motor,conderino2017linkage}, crash-to-emergency-service linkage shows the
same selection pattern \citep{hosseinzadeh2022data}, and an early descriptive study of pregnant
women in crashes established the characteristics such linkages recover \citep{weiss2002characteri}.

\begin{table}[htbp]
\rmfamily\centering
\small
\setlength{\tabcolsep}{4pt}
\caption{Literature positioning. No existing source covers a state population of police-reported crashes with a denominator and a misclassification adjustment.}
\label{tab:lit}
\begin{threeparttable}
\begin{tabular}{>{\raggedright\arraybackslash}p{2.1cm}>{\raggedright\arraybackslash}p{3.5cm}>{\raggedright\arraybackslash}p{2.4cm}>{\raggedright\arraybackslash}p{2.9cm}c>{\raggedright\arraybackslash}p{1.9cm}}
\toprule
\hdr{Study} & \hdr{Data source} & \hdr{Unit of analysis} & \hdr{Reported quantity} & \hdr{Features} & \hdr{Scope} \\
\midrule
\citet{viano2023} & NASS-CDS 2008--2015 and CISS 2017--2020 & towed passenger-vehicle crash & 2{,}467 $\pm$ 1{,}407 fetal deaths & F & national sample \\
\citet{chan2025} & level-I trauma center & injured patient & 157 injured patients & S F & one center \\
\citet{hattori2021} & NASS-CDS 2001--2015 & injured occupant & 736 injured women & S F & national sample \\
\citet{klinich2008} & crash investigation with reconstruction & pregnant occupant & case series, count not stated & S F & national sample \\
\citet{vladutiu2013aap} & birth and fetal-death records linked to crash reports & pregnant driver & cohort of 878{,}546 pregnancies & D S F & one state \\
\citet{vladutiu2013ajpm} & birth and fetal-death records linked to crash reports & pregnant person & cohort, count not stated & D S F & one state \\
\citet{hyde2003} & crash reports linked to birth and fetal-death records & pregnant person & 8{,}938 pregnant crash cases & D s F & one state \\
\citet{schiff2005} & hospitalization records linked to birth records & hospitalized patient & cohort, count not stated & D S F & one state \\
\citet{weiss2001} & fetal death certificates & fetal death & 240 traumatic fetal deaths, all mechanisms & F & 16 states \\
\citet{redelmeier2014} & self-matched driver cohort & pregnant driver & cohort of 507{,}262 women & D S & one province \\
\textbf{This paper} & police crash narratives and person-level records & police-reported crash & 5,333 confirmed crashes, 5,467 adjusted & D S F M & state population \\
\bottomrule
\end{tabular}
\begin{tablenotes}[flushleft]\footnotesize
\item[] Note: features are D for a denominator, S for gestational stage, F for a fetal outcome, and M for adjustment for classifier misclassification; a lower-case letter means partly available. Each figure is the one the study's own record gives, and the reported quantities are of different kinds, since a cohort size, an outcome-event count and a count of identified pregnant crash cases are not comparable. The column is a guide to scale rather than a like-for-like comparison.
\end{tablenotes}
\end{threeparttable}
\end{table}

\subsection{Police data, under-ascertainment, and narratives as a surveillance source}\label{sec:bg-police}
A count derived from police reports is a count of what officers recorded, and the gap between
that and what occurred is both large and patterned. Capture-recapture and linkage studies find
substantial under-ascertainment wherever they are run. Roughly two-thirds of Queensland
road-crash injuries had no matching police record \citep{watson2015estimating}, and French
national police counts scaled to 3.4 times their face value once registry-based correction was
applied \citep{amoros2008actual}. The underlying French linkage of 59{,}714 casualties put the
police reporting rate at 37.7\% and modeled it against crash and casualty characteristics
\citep{amoros2006underreporting}. In New Zealand, probabilistic linkage of hospital and police
records showed that police captured fewer than two-thirds of hospitalized vehicle-occupant crash
victims, with reporting varying by age, injury severity and length of stay
\citep{alsop2001underreporting}. Forty-five percent of Scottish road-casualty hospital admissions
were never recorded by police \citep{jeffrey2009an}, and Hong Kong reporting rates for crashes
requiring hospital care were 57.5 to 59.9\%, falling to about a third for children and cyclists
\citep{loo2007factors}. In Hyderabad only 2.3\% of non-fatal outpatient road-traffic injuries
reached the police \citep{dandona2008underrepor}, and across these settings under-reporting
concentrates among vulnerable road users
\citep{janstrup2016understand,macpherson2024comparison,tin2013completene}.

The methodological literature on measuring that gap is directly applicable here.
Capture-recapture estimates population size from overlapping incomplete ascertainment sources,
and the review by \citet{hook1995capture} sets out the two assumptions that fail most often,
namely independence between sources and homogeneous catchability across individuals. The
companion pair by the \citet{international1995capture,intlwg1995two} develops the log-linear
formulation and its applications to counting incident and prevalent cases where registration is
weak. \citet{tilling2001capture} argues that register-based estimates mislead when the registers
being compared were built for different purposes, which is exactly the situation when a police
file is compared against a clinical one. \citet{gibbons2014measuring} separates the two distinct
losses in the surveillance pyramid, under-ascertainment when a case never seeks care and
under-reporting when care is given but not notified, and compares the methods that correct each.
The evaluation framework of \citet{german2001updated} names sensitivity as a property of a
surveillance system rather than of a test, which is the sense in which this paper uses the term.
Our design departs from capture-recapture in one respect worth stating, since we have one
ascertainment source rather than two and therefore estimate the missing fraction from an external
expectation and from a direct audit of the unflagged stratum, rather than from the overlap
between lists.

Two findings from the police-data literature bear directly on the present design. First, what an
officer records depends on severity and circumstance rather than on the underlying condition,
since police documented alcohol involvement in only 72\% of drivers whose hospital-measured blood
alcohol exceeded the legal threshold, with documentation rising with severity, night-time and
single-vehicle crashes \citep{brubacher2013police}. Second, field-level completeness varies
systematically by attribute, so crash-database quality is not a single number
\citep{imprialou2019crash}. A variable that is not a coded field at all should be expected to
behave the same way, which is why this paper models the documentation process explicitly rather
than treating a narrative count as a prevalence estimate. The free-text narrative is where an
officer records what the form has no field for, and a substantial literature recovers information
from it. Rule-based methods and classical machine learning have identified work-zone crashes
misclassified in the coded fields \citep{sayed2021identifica}, agricultural crashes
\citep{kim2021crash}, occupational crashes through linkage to workers' compensation claims
\citep{northcutt2024}, contributory factors beyond the single code a form permits
\citep{zou2024crashcausi}, and latent themes across a decade of fatal-crash narratives
\citep{kwayu2021}. Deep learning extended this to pedal misapplication
\citep{bareiss2021finding} and to severity classification \citep{oliaee2023using,lee2023advancing}.
The injury-surveillance literature on auto-coding narratives is methodologically the same problem
and is more mature, including explicit human-machine ensemble designs and study of
rare-category behavior \citep{marucciwellm2017classifyin,bertke2016comparison,nanda2018improving}.
Two results in that literature are closest to ours. \citet{fitzpatrick2017an} used blind multi-reviewer reading of
604 narratives to validate how accurate a coded designation actually is, treating narratives as
the reference standard against which a coded field is judged, which is the inverse of the usual
arrangement. \citet{arteaga2025a} used a large language model over narratives specifically to
uncover under-reporting in coded fields, which is the use we make of them here. Language models
have since been applied to crash narratives directly. \citet{zhang2026} compared eleven models
across four paradigms on 16{,}656 Kentucky narratives for secondary-crash identification, and
\citet{su2026llmdriven} extracted structured causal frames from California narratives at 86.6\%
accuracy, rising to 97.9\% with sentence-level oversampling. \citet{crashsage2025} fine-tuned an
open model that outperforms tabular baselines on severity while exposing word-level risk factors,
\citet{chen2026narratives} predicted hazardous driver actions probabilistically, and
\citet{mumtarin2023llm} compared three public interfaces over 100 Iowa and Kansas narratives.
Police narratives outside the crash setting behave the same way, and \citet{relins2024vulnerability}
benchmarked instruction-tuned models against human coders on indicators of vulnerability in
police incident narratives while testing for coding bias.

\subsection{Validating model-extracted variables and adjusting for misclassification}\label{sec:bg-valid}
Because the extractor here is a model, the relevant methodological literature is the one on
validating model-extracted variables against human adjudication. In clinical text this is well
developed and the results are consistent, since agreement with expert labels is high for the
primary presence question and degrades on finer attributes. Zero-shot grading of postoperative
complications reached binary accuracy of 0.93 to 0.94 against a two-clinician benchmark
\citep{warmer2026zeroshot}, while extraction of adverse events from cell-therapy notes was
accurate for severity grading but only moderate for secondary attributes
\citep{guillot2026extracting}. Agreement with human raters has in several studies matched or
exceeded human inter-rater agreement \citep{kristjnsson2026prompting,li2026measuring}. Hallucination, where measured against
a physician reference, has been rare and low-impact \citep{van2026validating}, though note-level
phenotyping shows a characteristically over-inclusive profile \citep{hier2026notelevel}. The same
approach has been used for public-health surveillance from free text
\citep{algaradi2025automated,wang2026extracting}, for retrieval and diagnosis from the electronic
record \citep{ibrahim2026diagnostic}, and for medication-error classification, where binary
concordance of 96.0\% fell to 76.5\% on subcategorization \citep{krifors2026improving}. The
established pipelines this replaces were rule-based and were themselves validated component by
component \citep{savova2010ctakes}. Three cautions from this literature shape our design.
Performance is strongly sensitive to prompt construction, with identical tasks varying from an
$F_1$ of 0.854 to 0.965 by prompting strategy alone \citep{wu2026automated}. Prompt architecture
induces order and label biases that persist even when the model is told the options were
randomized, and that are removed only by aggregating across prompt variants
\citep{brucks2025prompt}. Generative models do not always win, since fine-tuned encoders
outperformed contemporary decoder-only models on overdose-death classification
\citep{funnell2026improving}, and even where a general model beats crowd annotators on ordinary
text \citep{gilardi2023chatgpt} or extracts structured records from scientific prose
\citep{dagdelen2024structured}, the output is a label rather than a probability. That last point
is the one that matters for a population count. A label carries no error rate, so a count
assembled from labels rests on an unmodeled assumption about error, and this paper makes that
assumption explicit and measurable instead.

The tools for doing so are old and well characterized. \citet{rogan1978} gave the estimator that
recovers true prevalence from apparent prevalence when sensitivity and specificity are known, and
its properties for a misclassified binary outcome have recently been reviewed by
\citet{edwards2026the}. \citet{walter1988estimation} set out the estimation of test error rates
and prevalence jointly from misclassified data, and \citet{brenner1993use} showed how a predictive
value can be used for the same correction. What a validation substudy can legitimately estimate
depends on how it was drawn, and \citet{fox2020common} and \citet{holcroft1999design} show that
sampling on the misclassified measure, as we do, identifies sensitivity and specificity within the
sampled stratum rather than predictive values for the whole corpus. \citet{marshall1990validation}
compares the indirect adjustment in common use against the more efficient direct method and
derives the variance of each. Intervals must widen to absorb the fact that sensitivity and
specificity are themselves estimated \citep{lang2014confidence,reiczigel2010exact}, and
\citet{diggle2011estimating} gives closed-form corrections when the two are known and a Bayesian
interval when they are not. The regime in which the estimator breaks down is equally well
documented. \citet{sempos2021adjusting} show that at true prevalence of one to five percent assay
error produces a background of bias that can exceed the prevalence being estimated, and
\citet{gelman2020bayesian} show that prevalence estimates are then dominated by uncertainty in
specificity. \citet{flor2020comparison} show truncation of the point estimate at zero or unity
with substantial under-coverage of conventional intervals, and \citet{liu2023quantitati} gives
indicators for deciding when a correction is worth making at all. Where the bias parameters are
unknown rather than estimated, quantitative bias analysis reconstructs the data as they would
have been under stated assumptions \citep{lash2003semiautoma,lash2009applying,lash2014good}.
\citet{greenland2005multiplebias} extends that practice to several biases at once, and
\citet{greenland2009bayesian} and \citet{maclehose2012is} connect it to the Bayesian treatment
of the same problem, the two performing comparably outside unrealistic priors. The analysis
below uses each of these in the place its assumptions hold, and the county smoothing and trend
models draw on the standard small-area and crash-count literature
\citep{clayton1987empirical,marshall1991mapping,hauer1997observational,lord2010statistical}.

No existing study has done all five of the following at once. None has used police narratives to
identify pregnancy at state-population scale, reported road-user role, gestational stage and
post-crash condition including documented fetal harm, adjusted the counts for measured classifier
misclassification, modeled the documentation process itself, and supplied denominators from
person-level crash records. Each piece exists somewhere in the literature above, and the
combination is what turns a narrative count into a surveillance estimate with a stated error
structure. This paper reports that combination for one state across nine years.

\section{Data and methods}\label{sec:methods}

\begin{figure}[tbp]\centering
\includegraphics[width=\textwidth]{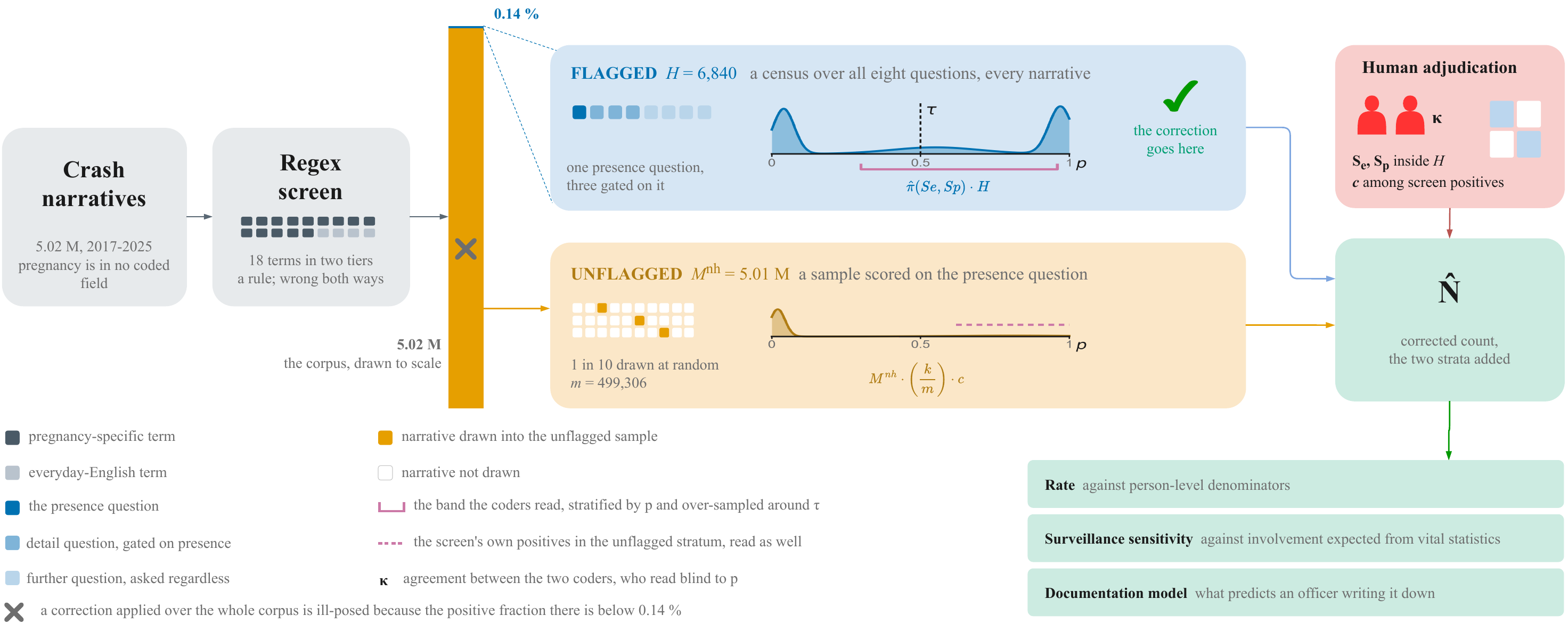}
\caption{Design of the study. A regular-expression screen splits the corpus into a flagged
stratum and a much larger unflagged remainder, drawn to scale, so the flagged fraction is the
hairline that the callout opens out into the upper band. The two strata are measured with
different instruments, a census of eight typed questions on one side and a one-in-ten random
sample scored on the presence question on the other, and the results are added. Coders blind to
the model's probability supply sensitivity and specificity inside the flagged stratum. The
correction for classifier error is applied there, marked by the tick, and not over the corpus,
marked by the cross, where the apparent positive fraction is too near zero for that correction to
be well conditioned. Densities and token strips are schematic.}
\label{fig:framework}\end{figure}

Figure~\ref{fig:framework} sets out the design, and one ratio explains every choice in it. The screen flags a fraction of a percent of the corpus, so the flagged and unflagged parts of
the corpus cannot be measured with the same instrument, and the correction for classifier error
cannot be applied to the corpus as a whole. The flagged part receives a census of eight typed
questions and a human-adjudicated audit of the answers. The unflagged part receives a single
presence question asked of a large random sample, because a census there would be ten times the work and would answer a question about
recall that a sample answers just as well. The two
parts are then added back together to give a statewide count. Table~\ref{tab:notation} fixes the
notation that the rest of this section and Appendix~\ref{app:derivations} use.

\begin{table}[htbp]
\rmfamily\centering
\small
\caption{Notation used in the estimation chain.}
\label{tab:notation}
\begin{tabular}{l>{\raggedright\arraybackslash}p{10.0cm}}
\toprule
\hdr{Symbol} & \hdr{Meaning} \\
\midrule
$M_t$ & Narratives filed in year $t$ after the length filter \\
$H_t$ & Stage-A regex hits in year $t$ \\
$M_t^{\mathrm{nh}}$ & Non-hit narratives in year $t$, equal to $M_t-H_t$ \\
$O_t$ & Narratives among the $H_t$ hits the model calls positive at $\tau$ \\
$\tau$ & Decision threshold on the presence probability, fixed at 0.5 \\
$p_i$ & Model probability that narrative $i$ documents a pregnant person \\
$\hat q_t$ & Apparent prevalence within the hits, equal to $O_t/H_t$ \\
$\hat\pi_t$ & Misclassification-corrected prevalence within the hits \\
$\mathrm{Se},\ \mathrm{Sp}$ & Sensitivity and specificity of the extractor inside the hit stratum \\
$J$ & Youden index of the extractor, equal to $\mathrm{Se}+\mathrm{Sp}-1$ \\
$m,\ k$ & Non-hit narratives screened in Stage C, and model positives among them \\
$c$ & Share of Stage-C positives confirmed as true misses on adjudication \\
$\hat N_t$ & Adjusted count of crashes with a documented pregnancy in year $t$ \\
$\phi_t^{49}$ & Share of confirmed cases that are driver-role and match a female driver aged 15 to 49 \\
$\phi_t^{44}$ & The same share on ages 15 to 44, used against the fertility-rate denominator \\
$D_t$ & Female drivers aged 15 to 49 involved in crashes in year $t$ \\
$D_t^{44}$ & Female drivers aged 15 to 44 involved in crashes in year $t$ \\
$\rho_t$ & Point prevalence of pregnancy among women aged 15 to 44 \\
$E_t$ & Pregnant drivers expected in year $t$ \\
$S_t$ & Surveillance sensitivity, the adjusted count over the expected count \\
$\delta(x)$ & Probability that a true pregnancy is documented at covariates $x$ \\
$Y_i$ & Indicator that driver $i$ is pregnant and that it was recorded \\
$\pi_j$ & Inclusion probability of a narrative drawn in validation stratum $j$ \\
$w_i$ & Design weight of validation narrative $i$, equal to $1/\pi_j$ \\
$\lambda_g$ & Documentation rate of county $g$ \\
$\alpha,\ \beta$ & Gamma prior parameters of the county smoothing model \\
\bottomrule
\end{tabular}
\end{table}

\subsection{Data sources}\label{sec:data}
The Texas Crash Records Information System holds a free-text narrative written by the
investigating officer for every police-reported crash, and the extract used here covers crashes
from 2017 through 2025. That extract as supplied holds \nRowsRead{} narrative rows. Dropping \nExcludedEmpty{} empty narratives and
\nExcludedShort{} of \minChars{} characters or fewer leaves \nNarratives{}. Running the
Stage-A vocabulary over the excluded short rows matches \nShortHits{} of them, so the
length cut costs no detectable case, and the target population is the corpus of longer
narratives. Names and telephone numbers arrive replaced by placeholders,
and we apply a second regular-expression pass over dates, driver-license and identification
numbers, long digit runs, case numbers, titled names and dates of birth to every copy that a
human or an external service ever sees.

That second pass is not sufficient, and we measure what survives it rather than assuming the
text is clean. Screening the documented fetal-harm narratives with a residual-identifier
question found that \nPIIExcluded{} of \nPIIScreened{} narratives, which is
\pctPIIExcluded{}\%, still carried a literal name, address or identifier after both passes.
Those narratives are excluded from any display, and no narrative appears anywhere in this paper
in any form. The same screen over the \nPIIReviewScreened{} narratives of the adjudication
sample flags \pctPIIReviewFlagged{}\% of them, or \nPIIReviewFlagged{} rows, a rate consistent
with the fetal-harm series and a property of the supplied corpus rather than of our sampling.
Those rows are flagged for handling rather than deleted from the sample. The validation sample
is stratified on the model's own probability and weighted back to the population it was drawn
from, and narrative length is associated with both identifier count and pregnancy detail.
Deleting rows on that criterion would therefore bias the very sensitivity and specificity the
sample exists to estimate. The adjudication file is instead handled as a limited disclosure to
named coders under the data agreement rather than as a de-identified artifact.

A second extract, holding one record per person, supplies the denominators and the covariates.
Inspection at five offsets spanning the study period establishes that this extract holds one row
per unit, meaning the driver, or a
pedestrian or cyclist as a unit in their own right, rather than one row per person. The person-number field is one in every populated row, and among the female persons aged
15 to 49 this study draws its denominators from, \pctFemaleDrivers{}\% are drivers, so
passengers are absent from the extract and no passenger denominator can be built from it. That property governs the rate definitions of Section~\ref{sec:denom} and is the reason
passenger cases below are reported as counts. The extract yields \nFemaleDrivers{} unique female
persons aged 15 to 49 across the nine years, together with restraint, airbag, ejection, injury
severity, vehicle, roadway and county fields joined to each case by crash identifier.

Expected pregnancy counts require the point prevalence of pregnancy among Texas women of
reproductive age, and rather than reconstruct a population denominator we use an identity that
falls out of the published general fertility rate. Texas resident live births come from state
vital statistics through 2022 and from national final and provisional natality reports for 2023
and 2024, and state fertility rates come from the final natality report of each anchor year. The
resulting point prevalence runs from \prevPregLo{}\% to \prevPregHi{}\% across the study period.
Every source and access date is recorded in the released \texttt{external\_vitals.json}.

\subsection{Case ascertainment and extraction}\label{sec:ascertain}
Ascertainment runs in three stages, shown as a funnel in Figure~\ref{fig:funnel} and counted in
Table~\ref{tab:funnel}.

Stage A is a regular-expression screen applied to every narrative, term by term, so that
per-term and per-term-unique hit counts are recoverable. The expression has two tiers, listed in
full with their counts in Table~\ref{tab:regex}. A core tier holds pregnancy-specific vocabulary
such as \texttt{pregnan}, \texttt{unborn}, \texttt{fetus}, \texttt{trimester}, \texttt{placent},
\texttt{gestation} and \texttt{miscarr}, and an everyday-English tier holds words that appear in
obstetric contexts but far more often do not, namely \texttt{expecting}, \texttt{with child},
\texttt{contractions} and \texttt{due date}. The core tier hits \nHitsCore{} narratives, the
everyday-English tier adds \nHitsNoisyOnly{} that no core term catches, and the two together hit
\nHitsExpanded{} narratives, which is \prefilterPrevalence{}\% of the corpus. A narrower
core-only vocabulary yields \nHitsNarrow{} hits, so the everyday-English tier is what recovers
the remainder. The tier is worth including precisely because the model rather than the screen
does the discrimination, and the per-term counts are released so that the contribution of each
term can be audited.

Stage B sends all \nHitsExpanded{} hits to Jev with the eight-question \texttt{pregnancy\_v1}
schema, reproduced in full in Table~\ref{tab:schema}. The schema asks one presence question and
three gated choice questions covering road-user role, post-crash condition and gestational
stage. Four further questions cover transport to a medical facility, a driver medical episode
before or during the crash, narrative injury severity and whether a witness is cited.
A gated choice is forced to its \texttt{no\_pregnancy} option whenever the presence probability
is at or below the threshold $\tau=0.5$, and the ungated answer is retained for the sensitivity
analysis. The model returns a probability over the analyst-defined options and generates no
text, so every answer is a typed value with a number attached rather than a sentence that must
itself be parsed.

Stage C estimates the cases that the Stage-A screen failed to flag at all. A single lean presence
question, worded identically to the Stage-B presence question and carrying criteria of seven and
thirteen words, is asked of a large random sample of narratives the screen did not hit. A separate screening run over the same
corpus had already put that identical question to \nStageCScreened{} randomly sampled non-hit
narratives, so those answers are reused rather than asked a second time. The one respect in
which the earlier run differs is call composition, because it asks the question alongside nine
others rather than alone, and a model could in principle answer differently in company. We
measured that rather than assume it away. On a stratified paired set of \guardPairs{} narratives,
consisting of every available record above $p=0.2$ plus a random draw of clear negatives, the
question was asked alone and compared against the answer given in company. Binary agreement at
$\tau=0.5$ was \guardAgreement{}\%, the Pearson correlation was \guardR{}, and the mean absolute
difference in probability was \guardMAD{}. The reuse rests on that measurement rather than on an assumption.

\subsection{Validation and misclassification adjustment}\label{sec:val}
The validation sample is drawn from the Stage-A hits and stratified by the model's own
probability, with the ambiguous band $0.30 < p \le 0.95$ heavily over-sampled and random draws
taken from each tail. Every Stage-C positive and borderline narrative is adjudicated as a census
of its own population. Table~\ref{tab:weights} gives the stratum sizes, the draws and the
resulting inclusion probabilities. The design deliberately over-represents the band where the
model is least certain, so every rate below is weighted by the reciprocal of the inclusion
probability and carried back to the hit population. The unweighted value is reported beside it in
Table~\ref{tab:confusion} to show the size of the gap. Which bias parameters a validation
substudy can legitimately estimate depends on how it was sampled, whether on the gold standard,
on the misclassified measure, or at random. Sampling on the misclassified measure, as we do here,
identifies sensitivity and specificity within the sampled stratum rather than predictive values
for the whole corpus \citep{fox2020common,holcroft1999design}. That is exactly why the correction
below is applied within the hit stratum. Where sensitivity and specificity are themselves
estimated rather than known, the resulting prevalence interval must widen to absorb that
uncertainty \citep{lang2014confidence,reiczigel2010exact}, and the bootstrap described below does
so by resampling the validation set within stratum on every draw.

The labels against which the extractor is measured come from \labelSource{} rather than from any
model. Every narrative was read blind, meaning narrative text only,
with the model's probabilities withheld and presentation order shuffled, by coders working from
the written criteria of Table~\ref{tab:schema}. Two coders independently read the same
\nDoubleRead{} narratives, giving a Cohen's $\kappa$ of \kappaVal{} over the \nKappaPairs{} of
those that both called definitively, and a third coder read the remaining \nSingleCoded{}
narratives, which are Stage-C candidates added after the non-hit screen reached full coverage.
Those two groups together are the \nNarrativesRead{} narratives that were read. They are more
than the \nValStageA{} Stage-A and \nValStageC{} Stage-C narratives that enter the rates below,
because \nUnclearSetAside{} rows were marked uncertain by their reader and are set aside rather
than forced into a label.
Of the \nConflicts{} differences between the two double-coding readers, most are one coder
recording uncertainty where the other committed, and outright opposite calls number five. Every
one of those five falls in the probability band where the model is also unsure. They were settled
by the study lead against the written criteria, and the coders' original readings are retained,
so the agreement statistic reports what the two readers independently did rather than what was
decided afterward. Weighted sensitivity is \seVal{} with a 95\% interval of \seLo{} to \seHi{},
weighted specificity is \spVal{} with an interval of \spLo{} to \spHi{}, and the weighted
positive predictive value is \ppvVal{}, all from \nValStageA{} adjudicated Stage-A narratives.

Sensitivity and specificity are estimated by the Horvitz-Thompson ratio of weighted cell counts.
Write $w_i = 1/\pi_{j(i)}$ for the design weight of adjudicated narrative $i$ drawn in stratum
$j$, with $y_i$ for the adjudicated truth and $p_i$ for the model probability. The two estimators
are then
\begin{equation}\label{eq:ht}
\widehat{\mathrm{Se}} = \frac{\sum_{i}w_i\,\mathbb{1}\{y_i=1,\ p_i>\tau\}}{\sum_{i}w_i\,\mathbb{1}\{y_i=1\}},
\qquad
\widehat{\mathrm{Sp}} = \frac{\sum_{i}w_i\,\mathbb{1}\{y_i=0,\ p_i\le\tau\}}{\sum_{i}w_i\,\mathbb{1}\{y_i=0\}} .
\end{equation}
Both are ratios of weighted totals rather than simple means, so their variance is a ratio-estimator
variance and is evaluated by resampling within stratum rather than from a closed form.

Correction for classifier error uses the Rogan-Gladen estimator \citep{rogan1978}, whose
properties for a misclassified binary outcome are reviewed by \citet{edwards2026the}. Writing
$\hat q_t = O_t/H_t$ for the apparent prevalence among the $H_t$ hits of year $t$,
\begin{equation}\label{eq:rg}
\hat\pi_t \;=\; \frac{\hat q_t + \mathrm{Sp} - 1}{\mathrm{Se} + \mathrm{Sp} - 1},
\end{equation}
applied within the Stage-A hit stratum, where the adjudicated prevalence is \prevInHits{}
and the estimator is well conditioned. The quantity the estimator takes is the apparent
prevalence, which is the model-positive share of the hits, and the adjudicated figure is
quoted here because it is the one the validation measures. Applying it to the full corpus would be invalid rather than merely
imprecise. At a population prevalence near one in a thousand, any specificity short of near
perfection implies more false positives than true cases and the corrected count becomes
meaningless. That is the documented failure regime set out in Section~\ref{sec:bg-valid}, in
which the estimator truncates, the interval under-covers, and uncertainty in specificity
dominates everything else. Restricting the correction to a stratum whose apparent prevalence is
three orders of magnitude higher is what keeps it away from that regime.

What the screen missed is a different quantity on a different population and is estimated
separately, then added. Write $M_t^{\mathrm{nh}}$ for the non-hit narratives of year $t$, and
write $k$ and $m$ for the model positives and the narratives screened in Stage C. With $c$ for
the share of those positives confirmed on adjudication, the total for the year is
\begin{equation}\label{eq:total}
\hat N_t \;=\; \underbrace{\hat\pi_t\,H_t}_{\text{within the hits}}
\;+\; \underbrace{M_t^{\mathrm{nh}}\cdot\frac{k}{m}\cdot c}_{\text{what the screen missed}} .
\end{equation}
We pool $k/m$ across years rather than fitting it within each year, because prefilter recall is a
property of the regular expression against police writing rather than of the calendar. A per-year
rate built on a handful of positives also swings severalfold on a single chance observation. The
per-year variant is refitted as a sensitivity analysis and reported in Section~\ref{sec:res-val}.

Uncertainty travels through one bootstrap chain of \bootB{} draws rather than three separate
ones, because the same validation sample enters every quantity. Sensitivity and specificity are
carried through as one pair per replicate. Sensitivity is taken from the within-stratum
resampling of the validation set, one value per replicate, so the design weights survive
into the interval. Specificity cannot be, because the sample holds no false positive and
every resample returns one exactly, which would assert a specificity known without error.
It is drawn instead from the Jeffreys posterior on the adjudicated cells,
\begin{equation}\label{eq:beta}
\mathrm{Sp}^{(b)} \sim \mathrm{Beta}\!\left(n_{\mathrm{tn}} + \tfrac12,\; n_{\mathrm{fp}} + \tfrac12\right),
\end{equation}
which is what carries the uncertainty that zero errors in a finite sample warrants instead of
treating a boundary estimate as certain. Each draw then resamples the validation set within
stratum, recomputes Equation~(\ref{eq:ht}) from that resample, and resamples the Stage-C screen
together with the adjudicated share. Drawing the confirmation count binomially and evaluating
Equations~(\ref{eq:rg}) and (\ref{eq:total}) then gives one draw of the year's total.
\begin{equation}\label{eq:mc}
\hat N_t^{(b)} \;=\; \frac{O_t^{(b)}/H_t + \mathrm{Sp}^{(b)} - 1}{\mathrm{Se}^{(b)} + \mathrm{Sp}^{(b)} - 1}\,H_t
\;+\; M_t^{\mathrm{nh}}\,\frac{k^{(b)}}{m}\,c^{(b)},
\qquad b = 1,\dots,B .
\end{equation}
A delta-method variance, derived in Appendix~\ref{app:derivations} as Equation~(\ref{eq:delta}),
gives a closed-form check on which term dominates and shows that the sensitivity term carries the
interval whenever the corrected prevalence is far from zero. We report the observed count, the
probability-sum estimator $\sum_i p_i$ and the corrected count side by side, because the three
fail in different directions.

\subsection{Denominators, rates, and models}\label{sec:denom}
Because the person extract holds no passengers, rates are role-matched, meaning that cases whose
model-assigned role is driver are divided by female drivers aged 15 to 49 involved in crashes.
Passenger and non-occupant cases are reported as counts and carry no rate. This is a deviation
from a rate over all female occupants and it is the conservative reading, since a matched
numerator and denominator is the stricter comparison; what is lost is a passenger-specific rate.
The match is a condition on the numerator, and it excludes \nDriverUnmatched{} of the
\nRoleDriver{} driver-role cases, whose crash carries no female driver of that age in the
person extract. Narratives the screen missed carry no role at all, so their role
distribution is taken to match that of the confirmed ones. Writing $\phi_t^{49}$ for the
share of confirmed cases whose model-assigned role is driver and whose crash holds at least
one female driver aged 15 to 49 in the person extract,
and $D_t$ for female drivers aged 15 to
49,
\begin{equation}\label{eq:rate}
R_t \;=\; 1000\,\frac{\hat N_t\,\phi_t^{49}}{D_t} .
\end{equation}

Expected pregnant drivers are computed on ages 15 to 44, matching the denominator of the
published fertility rate, and $\phi_t^{44}$ is the same matched share on that narrower
band, so each comparison counts the numerator and the denominator over one population. A
general fertility rate averages fertility that varies sharply with age over every woman in
the band, and crash-involved drivers do not share that age distribution, so the expectation
is standardized to theirs. Writing $f_t(a)$ for the Texas age-specific birth rate in
five-year band $a$ and $D_t(a)$ for female drivers in that band, the expectation is
multiplied by $\sum_a D_t(a) f_t(a) / \big(\mathrm{GFR}_t \sum_a D_t(a)\big)$. Rates are
published for \ageStdYears{} years inside the study period and each report's own general fertility rate for Texas agrees with the value used
here to within \ageStdGfrGap{} per 1{,}000, which is how the right row was checked
rather than assumed. The factor runs from \ageStdLo{} to \ageStdHi{} and exceeds
one in every year, because crash-involved drivers are concentrated in the higher-fertility
ages, so the unstandardized expectation was too small and the surveillance figure built on
it was too large. Since the general fertility rate satisfies
$\mathrm{GFR}_t = 1000\,B_t/P_t$ for live births $B_t$ and female population $P_t$, the point
prevalence of pregnancy reduces to a quantity in which the population cancels,
\begin{equation}\label{eq:prev}
\rho_t \;=\; \frac{B_t\,(40/52)}{P_t} \;=\; \frac{40}{52}\cdot\frac{\mathrm{GFR}_t}{1000},
\end{equation}
where the leading factor is the share of a year that a full-term pregnancy occupies. The expected
count and the surveillance sensitivity follow as
\begin{equation}\label{eq:surv}
E_t \;=\; D_t^{44}\,\rho_t, \qquad S_t \;=\; \frac{\hat N_t\,\phi_t^{44}}{E_t} .
\end{equation}
Two assumptions attach to $S_t$ and only one of them signs the error. Live births exclude
pregnancies ending in loss or termination, so $\rho_t$ and $E_t$ are too small and $S_t$ is
too large on that account. The construction also assumes crash involvement is independent
of pregnancy, and Appendix~\ref{app:derivations} shows why the direction of that second
departure is not established by the literature this study can draw on. The reported figure
is therefore a live-birth scenario rather than a bound. Scenario prevalences for stated loss multipliers are in the released
outputs.

The documentation model asks which crashes carry a written-down pregnancy. Its eligible population is
female drivers aged 15 to 49 who are the only such person in their crash, and that restriction is
substantive rather than cosmetic. A narrative saying the driver was pregnant does not say which
driver when a two-vehicle crash has two female drivers of reproductive age, and attaching the
pregnancy to the wrong person would corrupt the very person-level covariates the model is about.
We drop \nAmbiguousDropped{} ambiguous case crashes from both sides, removing
\nAmbiguousPersons{} people, and keep only drivers the person file records as female and
aged 15 to 49. That leaves \nEligibleSole{} eligible drivers and \nDocCases{} documented
cases. Those cases are fewer than the \nRoleDriver{} confirmed driver-role narratives of
Table~\ref{tab:chars}, and the difference is not a single subtraction, because the three
restrictions are counted on different units. Table~\ref{tab:chars} counts every confirmed
driver-role narrative, while the model counts only those it can attach to one identified
female driver of reproductive age. We retain the permissive coding as a sensitivity
analysis. Writing $Y_i$ for documentation and $x_i$ for the covariate vector of driver $i$,
\begin{equation}\label{eq:doc}
\operatorname{logit}\Pr(Y_i = 1 \mid x_i) \;=\; \gamma_0 + x_i^{\top}\gamma .
\end{equation}
Controls are sampled at a known rate $f$, so the coefficients $\gamma$ are unbiased and only the
intercept is shifted, by $\log f$; it is corrected before any predicted probability is quoted.
Standard errors are cluster-robust by county, because documentation is an agency habit and
agencies map to counties far better than to individual crashes. The surveillance ratio would be the average of the documentation probability over the
pregnant population, in the sense of the identity
\begin{equation}\label{eq:identity}
S_t \;=\; \mathbb{E}\big[\,\delta(x)\,\big],
\end{equation}
where $\delta(x)=\Pr(Y=1\mid x)$ is the probability that a pregnancy is written down and
the expectation runs over the covariate distribution of pregnant occupants.
Appendix~\ref{app:derivations} gives the two lines behind it. Nothing in this paper estimates that expectation, and the
reason is worth stating plainly. The model of
Equation~(\ref{eq:doc}) is fitted on all eligible female drivers, whose outcome is the joint
event that a driver is pregnant and that the pregnancy was recorded. Its fitted probability
is therefore not $\delta(x)$, and its terms describe documented-case status rather than
officer behavior among pregnant drivers. Separating the
two would need pregnancy status for drivers whose pregnancy was never written down, which
no source in this study supplies.

Three further models complete the estimation chain described above. The annual matched driver count, which is the numerator of $R_t$, is modeled by a negative
binomial regression with a log exposure offset and an indicator for 2020,
\begin{equation}\label{eq:trend}
\hat N_t\phi_t^{49} \sim \mathrm{NB}(\mu_t,\alpha), \qquad
\log\mu_t = \beta_0 + \beta_1 (t - 2017) + \beta_2\,\mathbb{1}\{t = 2020\} + \log D_t ,
\end{equation}
with the negative binomial rather than the Poisson because the counts are corrected estimates
carrying their own uncertainty. Crash severity is modeled against documentation status by an
ordinal logistic regression over the five-level police severity scale, reported beside a
quantitative bias analysis. Because documented pregnancy is rare in the eligible population, the
observed odds ratio relates to the true one by a ratio of documentation probabilities,
\begin{equation}\label{eq:qba}
\mathrm{OR}_{\mathrm{obs}} \;\approx\; \frac{\delta_1}{\delta_0}\,\mathrm{OR}_{\mathrm{true}},
\end{equation}
where $\delta_1$ and $\delta_0$ are the probabilities that a true pregnancy is documented in a
serious and in a non-serious crash. Appendix~\ref{app:derivations} derives the approximation and
states when it holds. The bias analysis in the results then inverts that map. Over a grid of
assumed values of
$\delta_1$ and $\delta_0$ it reclassifies $d_s(1/\delta_s - 1)$ of the controls in severity
stratum $s$ into the case group, where $d_s$ is the documented count in that stratum, and refits,
\begin{equation}\label{eq:grid}
\mathrm{OR}_{\mathrm{adj}}(\delta_1,\delta_0)
= \frac{d_1/\delta_1 \big/ \big(N_1 - d_1/\delta_1\big)}{d_0/\delta_0 \big/ \big(N_0 - d_0/\delta_0\big)} ,
\end{equation}
with $N_s$ the population total of stratum $s$. This is the reconstruction described in
Section~\ref{sec:bg-valid}, and the implementation follows the released code of
\citet{fox2023}. County rates are smoothed by a gamma-Poisson
empirical-Bayes shrink toward the state rate, whose posterior mean and shrinkage weight are
\begin{equation}\label{eq:eb}
\hat\lambda_g = \frac{y_g + \alpha}{e_g + \beta}, \qquad
w_g = \frac{e_g}{e_g + \beta},
\end{equation}
for county $g$ with documented cases $y_g$ and exposure $e_g$, and with $\alpha$ and $\beta$
fixed by method of moments on the state rate and its excess variance. Table~\ref{tab:eqmap} maps
every equation above to the function, script and output file that implements it.

\section{Results}\label{sec:results}

\subsection{Ascertainment and validation}\label{sec:res-val}
\begin{figure}[tbp]\centering
\includegraphics[width=\textwidth]{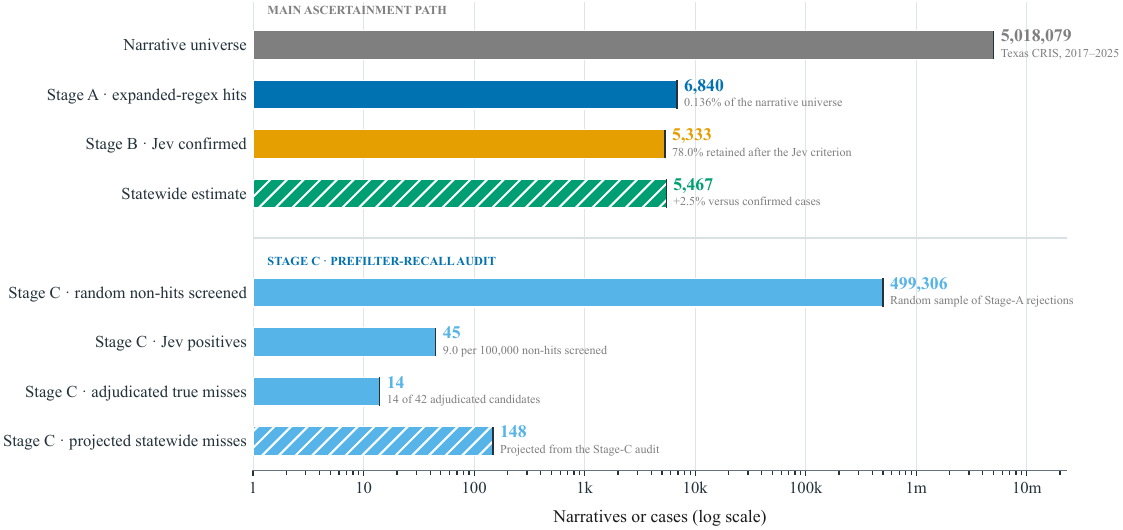}
\caption{Ascertainment funnel, with the Stage-C branch that estimates what the screen missed.
Counts are on a base-ten logarithmic scale, so bar lengths are not proportional to counts.
Hatched bars are estimates rather than observed counts, namely the statewide estimate and the
projected statewide misses.}
\label{fig:funnel}\end{figure}

Figure~\ref{fig:funnel} and Table~\ref{tab:funnel} give the funnel. Of the \nHitsExpanded{}
Stage-A hits, \nConfirmed{} are confirmed at $\tau=0.5$, and confirmation differs sharply by
tier. Core-term hits are confirmed at \coreConfirmRate{}\% against \noisyConfirmRate{}\% for
narratives matched only by everyday-English terms, so the everyday-English tier is almost
entirely false and earns its place only because the model rather than the screen decides which
of its hits survive. Table~\ref{tab:regex} shows where that asymmetry comes from. The single
term \texttt{pregnan} matches \nTermPregnan{} narratives and is the sole match for
\nTermSolePregnan{} of them, while \texttt{expecting} matches \nTermExpecting{} narratives and
\texttt{with child} \nTermWithChild{}, of which almost all carry no core term at all. Two core terms, \texttt{obstetric} and \texttt{foetus}, match nothing
in nine years of Texas police writing.

\begin{table}[htbp]
\rmfamily\centering
\small
\caption{Ascertainment funnel and validation of the extractor.}
\label{tab:funnel}
\begin{threeparttable}
\begin{tabular}{>{\raggedright\arraybackslash}p{8.2cm}>{\raggedleft\arraybackslash}p{4.2cm}}
\toprule
\hdr{Ascertainment step} & \hdr{Value} \\
\midrule
Narratives, 2017 to 2025, longer than 40 characters & 5,018,079 \\
Stage A, regex hits, core vocabulary only & 5,401 \\
Stage A, regex hits, expanded 18-term family & 6,840 \\
\quad of which matched a core pregnancy term & 5,416 \\
\quad of which matched an everyday-English term only & 1,424 \\
Stage B, sent to the model as a census of the hits & 6,840 \\
Stage B, confirmed at $\tau=0.5$ & 5,333 \\
Stage C, random non-hit narratives screened & 499,306 \\
Stage C, model positives among non-hits & 45 \\
Stage C, adjudicated candidates confirmed as true misses & 14 of 42 \\
Validation, Stage-A narratives adjudicated & 174 \\
Validation, Stage-C narratives adjudicated & 218 \\
Sensitivity, inverse-probability weighted & 0.999 [0.997, 1.000] \\
Specificity, inverse-probability weighted & 1.000 [1.000, 1.000] \\
Positive predictive value, inverse-probability weighted & 1.000 [1.000, 1.000] \\
Adjudicated prevalence within Stage-A hits, weighted & 0.782 \\
Cohen's $\kappa$ on the double-read narratives & 0.944 over 212 pairs \\
Misclassification-adjusted total, 2017 to 2025 & 5,467 [5,306, 5,577] \\
\bottomrule
\end{tabular}
\begin{tablenotes}[flushleft]\footnotesize
\item[] Note: sensitivity, specificity and predictive value are inverse-probability weighted to the Stage-A hit population, and each bracketed range is a 95\% bootstrap interval. Labels come from human adjudication. Coders read 407 narratives blind, 229 of them twice; 15 were marked uncertain and set aside, leaving the 174 Stage-A and 218 Stage-C narratives the rates are computed from.
\end{tablenotes}
\end{threeparttable}
\end{table}

\begin{figure}[tbp]\centering
\includegraphics[width=\textwidth]{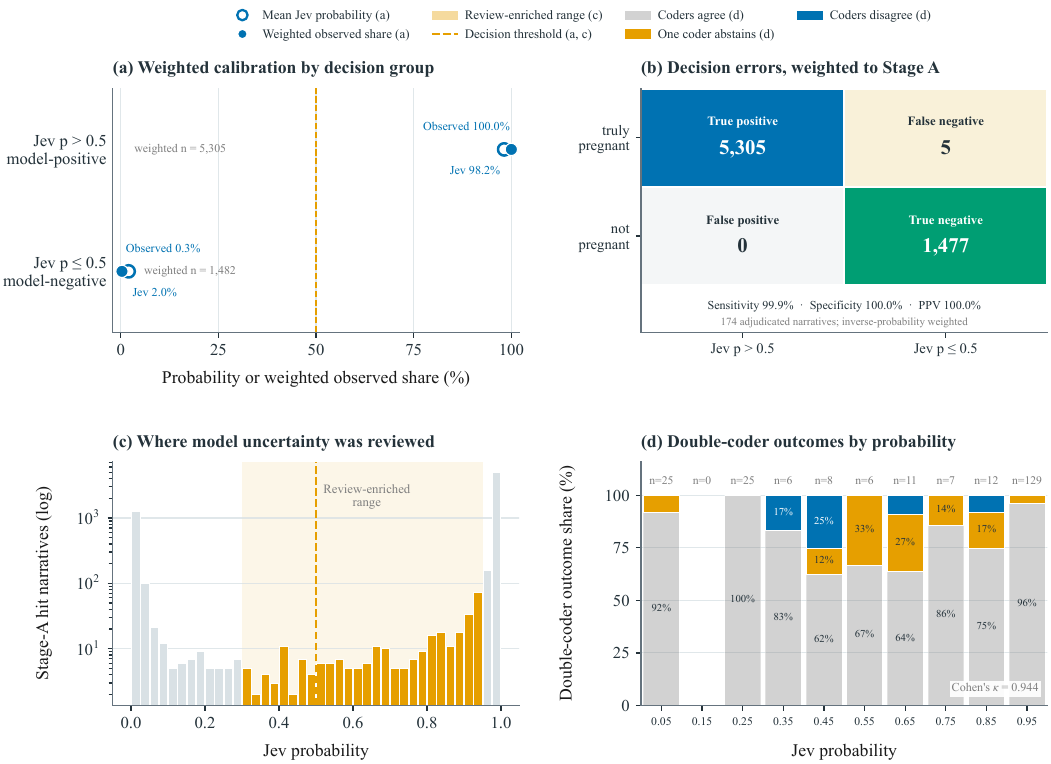}
\caption{Validation and calibration of the extractor. Panel (a) plots the mean model probability
as an open circle and the weighted share that adjudication confirmed as a filled circle, within
each decision group. Panel (b) gives the decision errors weighted to the Stage-A hit population.
Its four cells are drawn at equal size, so the block is a labeled contingency table rather than
an area-proportional mosaic. Panel (c) gives the distribution of model probabilities over
Stage-A hits on a logarithmic count axis, with the over-sampled review band shaded. Panel (d)
gives double-coder outcomes by probability bin.}
\label{fig:valid}\end{figure}

Figure~\ref{fig:valid} reports the validation of the extractor in four panels. Within each
decision group the model's mean probability is conservative against what adjudication found, as
panel (a) shows. Narratives the model calls positive are confirmed by adjudication more often than
its mean probability for that group states, and narratives it calls negative are confirmed less
often. In both directions the observed share is therefore further from the threshold than the
model stated, which is conservative rather than optimistic. Weighted to the hit population, the
decision errors of panel (b) are \cellFN{} false negatives and \cellFP{} false positives against
\cellTP{} true positives and \cellTN{} true negatives. Those cells give the sensitivity of \seVal{} and the specificity of
\spVal{} that Table~\ref{tab:confusion} sets out in both weighted and unweighted form. Panel (c)
shows why the sample was drawn the way it was. The probability distribution over Stage-A hits is
strongly bimodal, with more than a thousand narratives piled near zero and roughly four times as
many piled near one, a difference the logarithmic count axis compresses. The reviewed band from
\bandLo{} to \bandHi{} is the thin middle between those two peaks, where most bins hold a handful of
narratives and the fullest holds a few dozen, against the thousands in the two end bins. Panel (d) shows that the two coders
agree in more than nine cases in ten in the lowest and the highest probability bins. Agreement
falls to roughly two in three in the middle bins, where one coder abstains in as much as a third
of cases. Coder disagreement
and model uncertainty therefore sit in the same place, which is the pattern a calibrated
probability should produce.

Adjudication also yields a taxonomy of what makes the hard cases hard, which we report because it
is the part of the method that transfers to any other state. The dominant failure mode is the
word \emph{expecting} used in its ordinary sense, as in a driver expecting traffic to stop. The
remaining modes, in order of frequency, are narratives that name a pregnant person who was not
involved in the crash, elliptical statements of stage such as a bare reference to being six
months, and pregnancy that is suspected rather than stated. Below those sit pregnant road users
who are not crash units, and, on one occasion, a pregnant cow. Every one of these is a reading problem rather than a vocabulary problem, which is the
argument for a typed question over a longer word list.

Stage C screened \nStageCScreened{} random non-hit narratives and found \nStageCPositives{}
positives, a rate of \stageCRate{} per 100{,}000. Of the \nStageCReviewed{} candidates
adjudicated, \nStageCConfirmed{} were confirmed as genuine screen misses, which implies roughly
\nEstMisses{} missed cases statewide across the study period and a prefilter recall of
\prefilterRecall{}\%. The headline estimate pools that positive rate across years, and refitting
it within each year instead gives \nEstMissesPerYear{} missed cases against the pooled
\nEstMisses{}, a difference of well under one year's confirmations. The two agree because
prefilter recall is a property of the regular expression rather than of the calendar, which is
the assumption that pooling encodes. The per-year variant remains a check rather than an
independent estimate, since it rests on roughly \nPosPerYear{} Stage-C positives per year, thin
enough that a single chance positive moves a year's miss term substantially.

Role and documented outcome carry a reference standard built the same way as the
presence standard. Three coders read a random
\nRoleRead{} of the driver-role confirmations and all \nFetalRead{} fetal-harm
candidates, every row twice, with the reader who had not seen a row adjudicating it when
the other two differed. Agreement before adjudication was \roleAgree{}\% on role and
\fetalAgree{}\% on outcome. Of the \nRoleSettled{} rows whose role a coder could settle,
the model's driver label was confirmed in \nRoleConfirmed{}, and of the
\nFetalSettled{} settled outcomes its fetal-harm label was confirmed in
\nFetalConfirmed{}. Rows a coder marked unclear are excluded rather than forced, so these
are agreement among resolvable cases. The sample carries a second reading as well, since every row in it is a narrative the
model called pregnant. Of the \nRolePresenceSettled{} whose pregnancy a coder could
settle, \nRolePresenceConfirmed{} were confirmed, an unweighted precision of
\rolePPV{}\% beside the \ppvVal{} the probability-weighted sample gives. One row
divided its two readers and its adjudicator, and was settled by re-reading it blind with
all three coders, who agreed. A single adjudicator is not a reference standard where the
reading is contested, and the rule and the superseded answer are both in the released
validation record. Stage, transport and the remaining outcome
categories carry no reference standard, so those stay model outputs whose error rates are
unknown.

Three assumptions behind those rates are testable and were tested. The chain applies one
screen-miss rate to every year, and a chi-square test of homogeneity over the nine annual
Stage-C screens gives $\chi^{2}=\screenChi{}$ on \screenChiDf{} degrees of freedom with
$p=\screenChiP{}$, so the years are consistent with one rate. The \nUncertainLabels{}
adjudications a coder could not settle are excluded from the rates, and assigning every one
of them to the answer that hurts most moves sensitivity no lower than \seWorst{} and
specificity no lower than \spWorst{}. The Stage-C term assumes the model finds every pregnancy the screen missed, and that
assumption was tested where it can fail. Drawing at random from the unflagged stratum
cannot test it, because the quantity is too rare for any affordable sample to see, so the
reading was concentrated on the narratives the model came closest to flagging. All
\nStageCCensus{} it scored at \stageCCensusLo{} or above were read blind, together with
samples of the two bands below and the earlier random draw, \nStageCReadAll{} narratives
in all. The number describing a pregnancy is \nStageCFoundAll{}, and \nStageCUnsettled{}
could not be settled. Over the region where the model was at all hesitant, \nStageCTestedPool{}
narratives of which \nStageCTestedRead{} were read and two bands were read entire, the
95\% upper bound is \stageCUpperStratum{} further cases in the screened sample, or about
\stageCUpperState{} statewide against a total of \nAdjustedTotal{}. Below that region every narrative carries the model's smallest reported value, so there is
no gradient to sample along and a random draw of any affordable size bounds nothing. Those
\nFloorBand{} narratives were therefore searched in full for any of \nSecondOrderTerms{}
further ways of naming a pregnancy that the screen does not carry, among them maternity,
prenatal, postpartum, in utero, gravid, amniotic, midwife, stillborn and caesarean.
\nFloorFrame{} matched. Both were read blind by all three coders, and \nFloorPregnant{}
states a pregnancy. What that bounds is a conjunction rather than a stratum, since a case
missed there is one written with none of the screen's terms and none of these. As a quantitative backstop, halving the model's sensitivity in the unflagged stratum
raises the miss term from \missTerm{} to \missAtHalf{} and the statewide total to
\totalAtHalf{}.

\subsection{Counts, rates, trend, and surveillance sensitivity}\label{sec:res-counts}
\begin{figure}[tbp]\centering
\includegraphics[width=\textwidth]{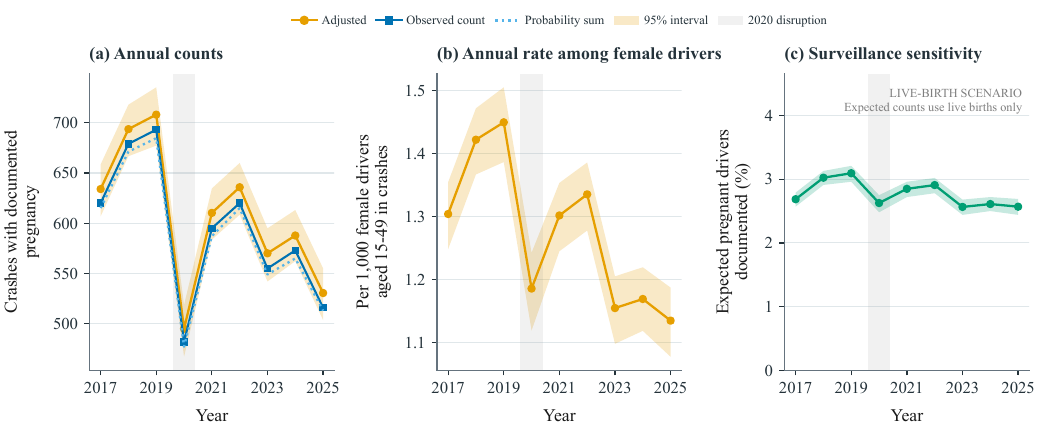}
\caption{Annual series, 2017 to 2025. Panel (a) gives counts, observed and corrected for
classifier error, with the probability sum shown for comparison. Panel (b) gives the corrected
rate per 1{,}000 female drivers aged 15 to 49 involved in crashes. Panel (c) gives the share of
expected pregnant drivers that narratives document, computed against an expectation built
from live births alone, which omits pregnancies ending otherwise. Bands are 95\% intervals and 2020 is shaded.}
\label{fig:annual}\end{figure}

Table~\ref{tab:annual} and Figure~\ref{fig:annual} give the annual series. The corrected statewide total for 2017 through 2025 is \nAdjustedTotal{} crashes with a
documented pregnancy, with a 95\% interval of \nAdjustedLo{} to \nAdjustedHi{}, against \nConfirmed{} raw confirmations and a
probability-sum estimate of \nProbSum{}. The three estimators agree closely, which is what a
specificity of \spVal{} implies, and panel (a) of Figure~\ref{fig:annual} shows the corrected
series lying a little above the observed series in every year while the probability sum sits at
or just below it. The shape of the series is the shape of Texas traffic over the period. Counts rise to a
maximum of \nCountMax{} in \yearCountMax{} and fall to \nCountMin{} in \yearCountMin{}. They
then recover over the two years that follow, and move unevenly downward across the last three,
ending below where the series began.

\begin{table}[htbp]
\rmfamily\centering
\small
\setlength{\tabcolsep}{3.5pt}
\caption{Annual estimates, rates, and surveillance sensitivity, 2017 to 2025.}
\label{tab:annual}
\begin{threeparttable}
\begin{tabular}{lrrrrrrr}
\toprule
\hdr{Year} & \hdr{Hits} & \hdr{Confirmed} & \hdr{Adjusted count\nl{}(95\% CI)} & \hdr{Female\nl{}drivers\nl{}15--49} & \hdr{Rate per 1{,}000\nl{}(95\% CI)} & \hdr{Expected\nl{}pregnant} & \hdr{Surveillance\nl{}sensitivity\nl{}\% (95\% CI)} \\
\midrule
2017 & 784 & 620 & 634 [607, 658] & 326,934 & 1.30 [1.25, 1.35] & 15,800 & 2.69 [2.57, 2.79] \\
2018 & 837 & 679 & 694 [667, 718] & 330,397 & 1.42 [1.37, 1.47] & 15,538 & 3.02 [2.91, 3.13] \\
2019 & 889 & 693 & 708 [677, 735] & 341,068 & 1.45 [1.39, 1.51] & 15,918 & 3.09 [2.96, 3.21] \\
2020 & 654 & 482 & 495 [467, 519] & 260,713 & 1.19 [1.12, 1.24] & 11,774 & 2.63 [2.48, 2.75] \\
2021 & 745 & 595 & 610 [584, 635] & 315,164 & 1.30 [1.25, 1.35] & 14,294 & 2.85 [2.72, 2.96] \\
2022 & 787 & 620 & 636 [609, 660] & 318,703 & 1.34 [1.28, 1.39] & 14,637 & 2.91 [2.78, 3.02] \\
2023 & 726 & 555 & 570 [542, 595] & 315,766 & 1.15 [1.10, 1.21] & 14,174 & 2.56 [2.44, 2.68] \\
2024 & 732 & 573 & 588 [562, 613] & 311,451 & 1.17 [1.12, 1.22] & 13,911 & 2.61 [2.50, 2.72] \\
2025 & 686 & 516 & 531 [504, 555] & 300,823 & 1.13 [1.08, 1.19] & 13,249 & 2.57 [2.44, 2.69] \\
\bottomrule
\end{tabular}
\begin{tablenotes}[flushleft]\footnotesize
\item[] Note: rates are driver-role cases over female drivers aged 15 to 49 in crashes, the only denominator the person extract supports. Expected counts and surveillance sensitivity use ages 15 to 44, matching the published fertility rate, and rest on live births alone, which omits pregnancies ending otherwise and makes these sensitivities too large on that account. Each adjusted count is that year's bootstrap median, so the column need not add to the study-period total.
\end{tablenotes}
\end{threeparttable}
\end{table}

Rates remove most of that exposure effect and leave a gentler decline. The rate per 1{,}000
female drivers aged 15 to 49 involved in crashes runs from \rateMin{} to \rateMax{} across the
period, reaching its maximum in \yearRateMax{} and its minimum in \yearRateMin{}. The three
lowest values of the nine fall in the last three years of the series. The negative binomial trend of Equation~(\ref{eq:trend}) is fitted to the matched driver
count that the rate itself is built from, with dispersion estimated at \trendAlpha{}
rather than assumed. It gives an annual rate ratio of \trendRR{}, a decline of about
\trendPct{}\% a year, with a 95\% interval of \trendRRLo{} to \trendRRHi{} and
$p=\trendP{}$. That interval is taken across the bootstrap chain, so it carries the
correction uncertainty as well as the regression's own. Dropping the partial final year
gives \trendRRComplete{} with an interval of \trendRRCompleteLo{} to
\trendRRCompleteHi{}, and the 2020 rate ratio is \covidRR{}. The 2020 term is small once exposure
is offset, which says that the sharp fall in counts that year is mostly a fall in the number of
female drivers in crashes rather than a change in how often pregnancy was written down.

The surveillance comparison is the result that reframes all of the above. Against a point
prevalence of pregnancy of \prevPregLo{}\% to \prevPregHi{}\% among Texas women aged 15 to 44,
the expected number of pregnant drivers involved in crashes runs between \expectedMin{} and
\expectedMax{} a year. The corrected documented count of drivers runs in the hundreds. Against that expectation police narratives document \survMin{}\% to \survMax{}\% of the
pregnant drivers, as panel (c) of Figure~\ref{fig:annual} shows. That series varies little across nine years, taking its maximum of \survMax{}\% in
\yearSurvMax{} and standing at \survLast{}\% in \yearLast{}, so the shortfall is a
stable property of the reporting system rather than a feature of any one year. The final
year is provisional on both sides, because its births are carried forward from provisional
2024 counts, its female population is interpolated, and the crash file holds \coverageLastYear{}\% of the mean annual
narrative count of the eight complete years. Dropping it leaves \totalComplete{} crashes with a documented pregnancy across 2017 to
2024, a rate of \rateMinComplete{} to \rateMaxComplete{} and a surveillance sensitivity of
\survMinComplete{}\% to \survMaxComplete{}\%. The figure is a scenario built on live births, and only one of its two departures from the
truth has a known direction. Expected counts rest on live births alone and so omit
pregnancies that end otherwise, which makes them too small and this share too large. The
second departure assumes crash involvement is independent of pregnancy, and the literature
reports a second-trimester increase in serious crashes rather than a ratio above one across
a whole pregnancy and all police-reported crashes, so its direction is not established here.

\subsection{Who, when, where, and what was documented}\label{sec:res-who}
\begin{figure}[tbp]\centering
\includegraphics[width=\textwidth]{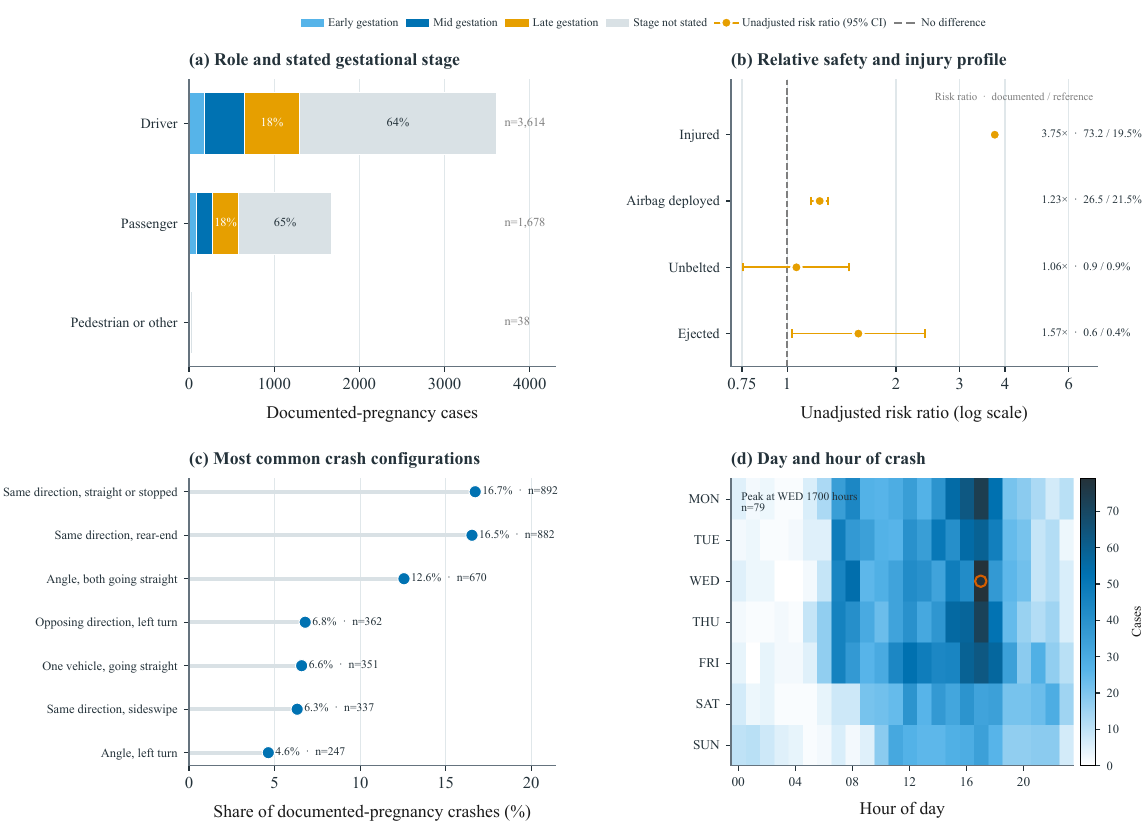}
\caption{Who, and how. Panel (a) gives road-user role and stated gestational stage. Panel (b)
gives unadjusted risk ratios for documented pregnant drivers against the eligible female drivers
of Table~\ref{tab:chars}, with both underlying percentages printed beside each ratio. Panel (c)
gives the most common crash configurations as a share of documented-pregnancy crashes. Panel (d)
gives the day and hour of crash, with the peak cell ringed.}
\label{fig:who}\end{figure}

Among confirmed cases, \pctRoleDriver{}\% are drivers, \pctRolePassenger{}\% are
passengers and \pctRolePedOther{}\% are pedestrians or other non-occupants. Three
confirmed cases carry no usable role and fall in none of the three categories. Gestational stage is stated in a
minority of narratives, with \pctStageNotStated{}\% giving no weeks, months or trimester against
\pctStageLate{}\% late, \pctStageMid{}\% mid and \pctStageEarly{}\% early pregnancy. Panel (a) of
Figure~\ref{fig:who} shows that drivers and passengers have almost identical stage composition,
at \pctLateDriver{}\% and \pctLatePassenger{}\% late and \pctNotStatedDriver{}\% and
\pctNotStatedPassenger{}\% not stated. Whether stage gets written down therefore
looks like a property of the officer's account rather than of the road-user's role.

Table~\ref{tab:chars} compares documented pregnant drivers with the eligible female drivers aged
15 to 49 against whom the documentation model is fitted, meaning those who are the only such
person in their crash. The differences are large and are mostly about detection rather than
risk. Documented
pregnant drivers are younger, at a mean age of \ageDoc{} against \ageAll{} years, and are
recorded as injured far more often, at \pctInjuredDoc{}\% against \pctInjuredAll{}\%, which
panel (b) of Figure~\ref{fig:who} expresses as an unadjusted risk ratio of \rrInjured{}. Airbag
deployment is higher at \pctAirbagDoc{}\% against \pctAirbagAll{}\%, a ratio of \rrAirbag{},
and ejection is higher at \pctEjectedDoc{}\% against \pctEjectedAll{}\%, a ratio of
\rrEjected{} whose interval only just clears one. Restraint use is the one clear exception to
that pattern. Unbelted status is \pctUnbeltedDoc{}\% against \pctUnbeltedAll{}\%, a ratio of
\rrUnbelted{} whose interval comfortably contains one, and it is one of the few
comparisons here not obviously distorted by the documentation mechanism. Crash configuration and
timing both look entirely ordinary. Same-direction and angle collisions account for the largest
shares, at \cfgPctOne{}\% for a same-direction crash with one vehicle straight or stopped,
\cfgPctTwo{}\% for a same-direction rear-end and \cfgPctThree{}\% for an angle crash with both
vehicles going straight. The
day-and-hour pattern peaks on a Wednesday at five in the afternoon.

\begin{table}[htbp]
\rmfamily\centering
\small
\caption{Documented pregnant drivers, and the eligible female drivers aged 15 to 49 that the documentation model draws its controls from.}
\label{tab:chars}
\begin{threeparttable}
\begin{tabular}{>{\raggedright\arraybackslash}p{5.6cm}>{\raggedleft\arraybackslash}p{3.2cm}>{\raggedleft\arraybackslash}p{3.4cm}}
\toprule
\hdr{Characteristic} & \hdr{Documented\nl{}pregnant drivers} & \hdr{Eligible female drivers\nl{}aged 15--49} \\
\midrule
Number of drivers & 3,614 & 250,000 \\
Mean age, years & 27.0 & 30.8 \\
Unbelted, \% & 0.9 & 0.9 \\
Airbag deployed, \% & 26.5 & 21.5 \\
Ejected, \% & 0.6 & 0.4 \\
Recorded as injured, \% & 73.2 & 19.5 \\
Serious or fatal crash, \% & 2.8 & 1.0 \\
Rural location, \% & 16.5 & 24.1 \\
Mean speed limit, mph & 43.1 & 42.2 \\
\bottomrule
\end{tabular}
\begin{tablenotes}[flushleft]\footnotesize
\item[] Note: the left column is every confirmed driver-role case, which is a wider set than the documented cases the documentation model is fitted on. The right column is that model's sampled control set, so it holds female drivers aged 15 to 49 who are the only such person in their crash, not all female drivers of that age. Its count is a sample rather than the state total; percentages are unaffected, because controls were drawn at a known constant rate.
\end{tablenotes}
\end{threeparttable}
\end{table}

\begin{figure}[tbp]\centering
\includegraphics[width=\textwidth]{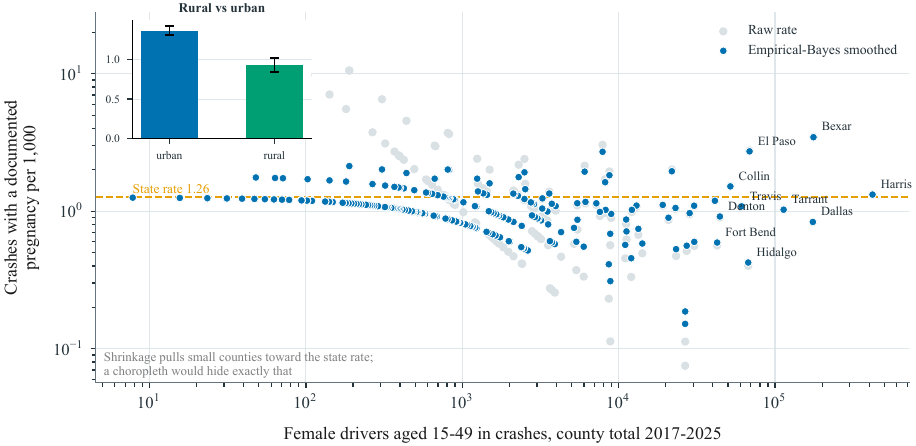}
\caption{County exposure against the empirical-Bayes smoothed documentation rate, with the ten
largest counties labeled and a rural and urban inset. Both axes of the main panel are
logarithmic. Pale points are raw rates and solid points are smoothed rates for the same county,
so the vertical distance between a pair is the shrinkage that county received.}
\label{fig:geo}\end{figure}

Geography behaves the way small-area counts usually behave, which is why
Figure~\ref{fig:geo} plots exposure against the smoothed rate rather than drawing a choropleth.
The empirical-Bayes shrink of Equation~(\ref{eq:eb}) pulls counties with few female drivers onto
the state rate of \stateCountyRate{} per 1{,}000 across \nCounties{} counties. The plot makes
that visible as a horizontal ridge of smoothed points at low exposure that fans out as exposure
grows. Among the ten largest counties the smoothed rates differ by a factor of \countyTopSpread{},
running from \countyHiRate{} per 1{,}000 in \countyHiName{} County down to \countyLoRate{} in
\countyLoName{} County, with \countyNearName{} County closest to the state rate among the ten at
\countyNearRate{}. The inset gives a
modest urban excess, at \rateUrban{} per 1{,}000 in urban counties against \rateRural{} in
rural ones, with intervals that do not overlap. Nothing in the county pattern tracks any plausible geography of pregnancy, and the spread
among large counties with similar populations is consistent with variation in documentation
practice. That reading is a hypothesis rather than a finding. Shrinkage controls random
instability and not case mix, the extract carries no agency identifier, and county
differences can also reflect age, injury pattern, crash type, urbanicity, access to care or
data completeness.

\begin{figure}[tbp]\centering
\includegraphics[width=\textwidth]{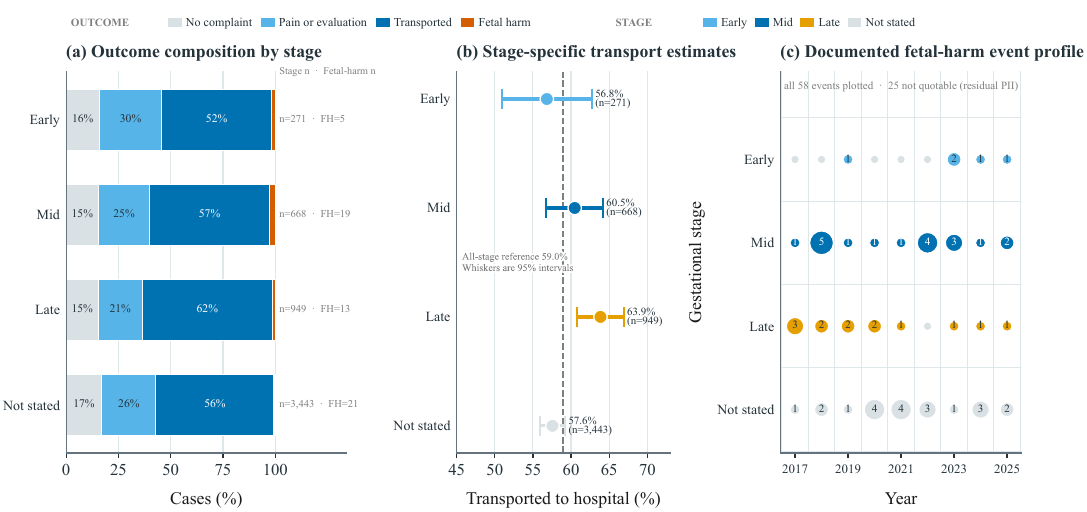}
\caption{Documented outcomes. Panel (a) gives outcome composition by stated gestational stage,
with the case count and the fetal-harm count printed beside each bar. Panel (b) gives the
stage-specific share transported to hospital against the all-stage reference, with 95\%
intervals. Panel (c) gives documented fetal-harm events by year and stage. All \nFetalHarm{} events are plotted; the panel's second count is the narratives that
residual identifiers keep from being quoted. The series counts fetal harm recorded in a
narrative rather than fetal harm caused by the crash, and adjudication found at least one
case in which the loss preceded the crash and explained it.}
\label{fig:outcomes}\end{figure}

Table~\ref{tab:outcomes} and Figure~\ref{fig:outcomes} give what the narratives record about the
pregnant person afterward. Among the confirmed cases that carry an outcome value, \pctOutTransported{}\% are
documented as transported to a hospital or medical facility, \pctOutPain{}\% as complaining of
pain or evaluated without transport, and \pctOutNoComplaint{}\% with no complaint recorded.
Transport rises with stated stage on both of the two measures the schema supports, and the two
ask about different people. Table~\ref{tab:outcomes} counts the gated post-crash condition,
which records what the narrative says about \emph{the pregnant person} and is mutually
exclusive across its four options. On that measure the transported share runs from
\nTransportedGatedEarly{} of \nStageEarly{} in early pregnancy to \nTransportedGatedLate{} of
\nStageLate{} in late pregnancy. Panel (b) of Figure~\ref{fig:outcomes} uses the standalone
transport question, which asks whether the narrative states that \emph{a person} was taken to a
hospital or medical facility and is asked of every narrative regardless of the gated answer. It
gives \transpEarly{}\%, \transpMid{}\% and \transpLate{}\% across the same three groups
against an all-stage reference of \transpAll{}\%. The two disagree in both directions rather
than one. On \nTranspStandaloneOnly{} narratives the standalone question records a transport
that the gated condition does not, and on \nTranspGatedOnly{} the gated condition records one
that the standalone question does not, which is what two questions about two different subjects
should be expected to do. On the standalone measure the interval for the late group lies entirely above the all-stage
reference while the intervals for the other three groups contain it. That comparison is
exploratory and is reported as a description rather than as a test. The reference contains
the groups compared against it, four groups are inspected with no global test and no
multiplicity adjustment, and both stage and transport are model-assigned fields that the
validation did not measure. The share complaining of pain without transport falls
across the same three groups, from \pctPainEarly{}\% to \pctPainMid{}\% to \pctPainLate{}\%,
while the share with no complaint stays near \pctNoComplaintLate{}\% throughout. That gradient is what a precautionary logic at the scene would produce, and it is the
clearest illustration of the documentation mechanism in the descriptive data. The further along a
pregnancy is stated to be, the more likely it is both that somebody mentions it at all and that
it changes what happens next.

\begin{table}[htbp]
\rmfamily\centering
\small
\caption{Documented outcome by stated stage of pregnancy.}
\label{tab:outcomes}
\begin{threeparttable}
\begin{tabular}{>{\raggedright\arraybackslash}p{2.6cm}rrrrrr}
\toprule
\hdr{Stated stage} & \hdr{No complaint} & \hdr{Pain or\nl{}evaluation} & \hdr{Transported} & \hdr{Fetal harm} & \hdr{Total} \\
\midrule
Early & 43 & 81 & 142 & 5 & 271 \\
Mid & 103 & 164 & 382 & 19 & 668 \\
Late & 146 & 199 & 591 & 13 & 949 \\
Not stated & 587 & 891 & 1,944 & 21 & 3,443 \\
All stages & 879 & 1,335 & 3,059 & 58 & 5,331 \\
\bottomrule
\end{tabular}
\begin{tablenotes}[flushleft]\footnotesize
\item[] Note: outcomes are what the narrative documents at the scene or in immediate follow-up, not clinical outcomes. Column totals run two below the confirmations reported elsewhere, because two confirmed narratives carry the gated no-pregnancy value. Of the 58 fetal-harm narratives, 25 (43\%) were withheld from display by the residual-identifier screen, which affects what can be shown rather than what is counted.
\end{tablenotes}
\end{threeparttable}
\end{table}

\nFetalHarm{} narratives, or \pctFetalHarm{}\% of the cases carrying an outcome value,
document harm to or loss of
the unborn child. We report that series only in aggregate and never case by case. Panel (c) of
Figure~\ref{fig:outcomes} spreads those events across nine years and four stage groups, and the
counts are in single digits everywhere. The events divide across stated stage as Table~\ref{tab:outcomes} records, and they show no
visible trend across years. No vignette and
no narrative text appears anywhere in this paper. Of the \nPIIScreened{} fetal-harm narratives,
\nPIIExcluded{} failed the residual-identifier screen outright, and the remainder are held
pending a further human clearance rather than paraphrased here.

\subsection{Documentation model and the severity association}\label{sec:res-doc}
\begin{figure}[tbp]\centering
\includegraphics[width=\textwidth]{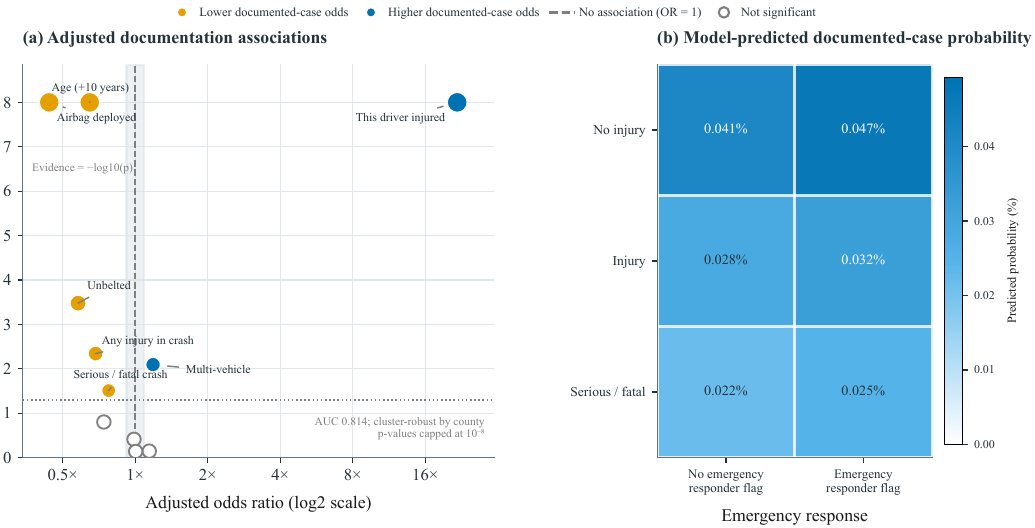}
\caption{Documentation model. Panel (a) plots adjusted odds ratios for being a documented
pregnant driver against the strength of evidence for each term, on a base-two odds-ratio
axis, with non-significant terms drawn hollow and $p$ values capped at $10^{-8}$. Panel (b)
gives the model-predicted probability of that joint outcome by crash-level injury severity
and emergency response, holding the driver's own recorded injury at its reference level.}
\label{fig:docmodel}\end{figure}

The documentation model of Equation~(\ref{eq:doc}) is fitted on a sample of \docN{} drivers across
\docClusters{} county clusters and reaches an area under the receiver operating characteristic
curve of \docAUC{}. Table~\ref{tab:docmodel} and Figure~\ref{fig:docmodel} give
its terms, and one of them dominates everything else. Whether this driver is herself recorded as
injured carries an odds ratio of \orInjured{}, with a 95\% interval of \orInjuredLo{} to
\orInjuredHi{}. That term sits alone at the right of panel (a), while every other term in the
model lies within a factor of two and a half of one. Documentation also falls steeply with driver
age, at \orAge{} per ten years with an interval of \orAgeLo{} to \orAgeHi{}, which we read as
mostly reflecting true pregnancy prevalence by age rather than officer behavior. The outcome is
the product of prevalence and documentation propensity, and the model cannot separate the two. Airbag
deployment carries an odds ratio of \orAirbag{} and unbelted status \orUnbelted{}, while the
presence of an
emergency responder, rural location, calendar year and speed limit are all indistinguishable from
one.

\begin{table}[htbp]
\rmfamily\centering
\small
\caption{Documentation model. Odds of being a documented pregnant driver, among female drivers aged 15 to 49 who are the only such person in their crash.}
\label{tab:docmodel}
\begin{threeparttable}
\begin{tabular}{>{\raggedright\arraybackslash}p{6.4cm}rrr}
\toprule
\hdr{Term} & \hdr{Odds ratio} & \hdr{95\% CI} & \hdr{$p$} \\
\midrule
Crash injured someone & 0.685 & [0.528, 0.890] & 0.005 \\
Serious or fatal crash & 0.776 & [0.617, 0.977] & 0.031 \\
This driver recorded as injured & 21.723 & [15.920, 29.643] & $<$0.001 \\
Emergency responder present & 1.144 & [0.553, 2.365] & 0.717 \\
Driver unbelted & 0.579 & [0.430, 0.781] & $<$0.001 \\
Airbag deployed & 0.440 & [0.397, 0.488] & $<$0.001 \\
Rural location & 0.741 & [0.489, 1.123] & 0.158 \\
Driver age, per ten years above 30 & 0.648 & [0.625, 0.671] & $<$0.001 \\
Calendar year, per year after 2021 & 0.989 & [0.963, 1.015] & 0.388 \\
More than one unit involved & 1.186 & [1.045, 1.346] & 0.008 \\
Speed limit, per 10 mph above 40 & 1.005 & [0.977, 1.034] & 0.729 \\
\bottomrule
\end{tabular}
\begin{tablenotes}[flushleft]\footnotesize
\item[] Note: logistic regression on 252,465 drivers, of whom 2,465 have a documented pregnancy. Standard errors are cluster-robust over 254 counties and the area under the receiver operating characteristic curve is 0.814. Controls were sampled at 0.128, which leaves odds ratios unaffected and shifts the intercept by the log of that rate; the intercept is corrected before any predicted probability is quoted.
\end{tablenotes}
\end{threeparttable}
\end{table}

The direction of the crash-level terms repays a careful reading, because it is the opposite of
what a casual glance would suggest. Conditional on whether this driver is recorded as injured,
every crash-level severity term points downward. A crash that injured somebody carries an odds
ratio of \orInjuredAny{} and a serious or fatal crash \orSerious{}. Panel (b) of Figure~\ref{fig:docmodel} traces
the consequence, with the predicted documentation probability falling from \docProbNoInjury{}\%
in a crash with no injury to \docProbSerious{}\% in a serious or fatal one. That probability
rises only slightly, by a few thousandths of a percentage point, when an emergency responder is
flagged. What the model therefore describes is a documented case that tracks the individual becoming
a medical subject rather than the crash becoming a severe event. The pattern is consistent
with a recording rule that fires when the driver herself is hurt, and that reading is an
interpretation rather than an estimated mechanism.

Equation~(\ref{eq:identity}) would connect this pattern to the surveillance number if the
fitted probability were the probability of documentation given pregnancy, and it is not.
The outcome is the joint event that a driver is pregnant and that the pregnancy was recorded, so
the predicted probabilities of panel (b) are a few hundredths of a percent and cannot
average to a surveillance sensitivity of a few percent. The two results are consistent in
direction and are reported side by side, and closing the arithmetic between them needs
pregnancy status for undocumented drivers.

Among eligible female drivers, documented pregnancy is associated with higher crash severity. The
ordinal model over the five-level severity scale gives an odds ratio of \sevOR{} with an interval
of \sevORLo{} to \sevORHi{}, and collapsing the outcome to serious or fatal gives \sevORBinary{}.
Refitting the binary model without the case-control population weights returns
\sevORBinaryUnw{} to three decimal places, so the weighting contributes none of the gap
between the two. The rest of it is the proportional-odds assumption failing. An ordinal
model asserts one exposure odds ratio at every cumulative threshold, and fitting the
\poCutN{} thresholds separately gives odds ratios from \poCutMin{} to \poCutMax{}, a
spread of \poSpread{}. Almost all of it sits at the threshold that separates any injury
from none, which is the same pattern the documentation model shows, so the ordinal
estimate is an average over thresholds that differ rather than one effect. We report those numbers only so that the next
one can be read beside them. Because documentation is itself severity-dependent, the undocumented
pregnancies are not missing at random. They sit in the control group, concentrated in exactly the
minor crashes where nobody had a reason to write pregnancy down, and Equation~(\ref{eq:qba}) says
that the observed association inherits the ratio of documentation probabilities between the two
severity strata. Reclassifying those pregnancies out of the controls stratum by stratum over a grid of
\biasGridN{} assumed documentation probabilities, running from \biasDocMin{} upward so
that the grid reaches the order of the surveillance estimate, moves the odds ratio across
\sevBiasLo{} to \sevBiasHi{}. Table~\ref{tab:severity} summarizes it and
Table~\ref{tab:biasgrid} and Figure~\ref{fig:bias} give it in full. That envelope is too
wide to bound anything, which is the result rather than a failure of it. What fixes the
sign is not the data but which stratum is better documented. Where serious crashes are documented better than minor ones, the odds ratio runs
from \biasSeriousLo{} to \biasSeriousHi{}, which holds the whole protective side of the
grid and reaches barely above one at its top. Where documentation is equal
in the two strata the grid returns \biasEqualLo{} to \biasEqualHi{}, which recovers the
naive estimate and is the check that the reclassification does what it claims. The envelope is computed on the binary outcome, because the
reclassification needs frequency weights, so it is the value of \sevORBinary{} that it should be
compared against, and the ordinal fit is reported as the conventional specification.

\begin{table}[htbp]
\rmfamily\centering
\small
\caption{Crash severity against pregnancy documentation, with the quantitative bias analysis.}
\label{tab:severity}
\begin{threeparttable}
\begin{tabular}{>{\raggedright\arraybackslash}p{7.4cm}rr}
\toprule
\hdr{Term} & \hdr{Odds ratio} & \hdr{95\% CI or range} \\
\midrule
Pregnancy documented, ordinal model & 4.149 & [3.876, 4.442] \\
Pregnancy documented, serious or fatal outcome & 1.622 & population weighted \\
Pregnancy documented, serious or fatal outcome, unweighted & 1.622 & unweighted \\
Detection-bias envelope on the binary outcome & 0.030 to 88.119 & 81 scenarios \\
Driver age, per ten years above 30 & 1.097 & [1.086, 1.107] \\
Driver unbelted & 5.910 & [5.433, 6.430] \\
Airbag deployed & 4.412 & [4.327, 4.499] \\
Rural location & 0.811 & [0.794, 0.828] \\
More than one unit involved & 1.310 & [1.273, 1.349] \\
Speed limit, per 10 mph above 40 & 1.087 & [1.081, 1.093] \\
Calendar year, per year after 2021 & 0.993 & [0.990, 0.997] \\
\bottomrule
\end{tabular}
\begin{tablenotes}[flushleft]\footnotesize
\item[] Note: the estimates are associations, not causal effects. Documentation is itself severity-dependent, as Table~\ref{tab:docmodel} shows, so the binary serious-or-fatal estimate carries the detection-bias envelope of the fourth row, which reclassifies undocumented pregnancies out of the control group stratum by stratum. The envelope is computed on that binary outcome and should be read against it rather than against the ordinal model of the first row. The full grid is Table~\ref{tab:biasgrid}.
\end{tablenotes}
\end{threeparttable}
\end{table}

The envelope spans one from end to end, and the no-association boundary runs through the
middle of the grid rather than along its edge. The apparent association between
pregnancy and crash severity in these data is therefore consistent with a substantial positive
association, with no association at all, and with a protective one, depending entirely on an
assumption that these data cannot test. We make no claim about whether pregnancy is associated
with crash severity, and we would discourage anyone from extracting one from narrative-derived
counts without the same analysis.

\section{Discussion}\label{sec:discussion}

The central finding of this study is that police narratives document a small fraction of
the pregnant road users who are involved in crashes, and that the crashes they document are
the ones in which the driver was hurt. Nine years of Texas narratives yield \nAdjustedTotal{} crashes with a
documented pregnancy and \nAdjustedDrivers{} with a documented pregnant driver, against an
expectation of \expectedTotal{} pregnant drivers over the same period. Those two nine-year totals stand in the same ratio that
Section~\ref{sec:res-counts} reports year by year, where the annual surveillance
sensitivities run from \survMin{}\% to \survMax{}\%. Placing that
number beside the existing estimates requires care about units, because the quantities differ.
Our unit is a police-reported crash in which an officer wrote that somebody involved was
pregnant. The unit of \citet{viano2023} is a towed passenger-vehicle crash in a weighted national
investigation sample, with an investigator's assessment of pregnancy. That study estimated
2{,}467 fetal deaths with an interval of plus or minus 1{,}407, across what its own conclusions
describe as a twelve-year sample.
Ours is far larger and covers the minor crashes that an investigation sample excludes, and theirs
has a clinical depth ours cannot approach. The linked-record studies give the comparison that
matters most, because they measure involvement rather than outcome. \citet{vladutiu2013aap}
estimated a crash risk of 12.6 per 1{,}000 pregnant women in North Carolina across 878{,}546
pregnancies, and \citet{hyde2003} identified 8{,}938 pregnant women in Utah crashes through
linkage. Our documented rate of \rateMin{} to \rateMax{} per 1{,}000 female drivers aged 15 to 49
involved in crashes is not the same quantity as either, because our denominator is crash-involved
drivers rather than pregnancies. The comparison that can be made is the one we make, against the
involvement that Texas vital statistics imply. A documented count smaller than expected
involvement by a factor of \survShortfall{} is the result. Any study treating narrative mentions
as a prevalence measure, or comparing narrative-derived counts across agencies or years without
modeling documentation, is therefore measuring officer behavior as much as epidemiology.

The series falls, and what falls is documentation. The rate reaches its maximum in
\yearRateMax{} and its minimum in \yearRateMin{}, and the three lowest of the nine values are
the last three years. A shortage of pregnancies does not explain it. The surveillance share
divides documented cases by an expectation built from Texas live births, and that share falls
over the same window, from \survMax{}\% in \yearSurvMax{} to \survMin{}\% in
\yearSurvMin{}, so the denominator moves the wrong way for that reading. What the series cannot
separate is a change in how often a pregnant woman is in a crash from a change in how often an
officer writes it down. Texas across this window is not a neutral setting for that distinction,
since state law bearing on pregnancy changed substantially during it, and both what an occupant
volunteers at the scene and what an officer records are plausibly sensitive to that. We make no
claim about which mechanism operates, because these data cannot support one. We raise it because
a declining series of documented pregnancy in this state over these years invites the question,
and leaving it unasked would be the weaker choice. Separating the two needs pregnancy counts from
outside the crash record, which is the linkage study Section~\ref{sec:conclusion} argues for.

That shortfall is the same phenomenon the under-ascertainment literature has measured for other
variables, and it sits at the severe end of the range that literature reports. Police captured
fewer than two-thirds of hospitalized vehicle-occupant crash victims in New Zealand
\citep{alsop2001underreporting}, and French counts scaled by a factor of 3.4 once registry
correction was applied \citep{amoros2008actual}, with a reporting rate of 37.7\% in the
underlying linkage \citep{amoros2006underreporting}. Roughly two-thirds of Queensland crash
injuries had no police record \citep{watson2015estimating}, and 45\% of Scottish casualty
admissions were never recorded \citep{jeffrey2009an}. Those figures describe casualties that the police did not record at all.
Our figure describes a specific attribute of a person the police did record, which is a different
and in principle easier task, and the attribute is nonetheless captured for only a few percent of
those who have it. The reason is the one \citet{brubacher2013police} found for alcohol, where
police documented involvement for 72\% of drivers whose measured blood alcohol exceeded the
threshold and documentation rose with severity. Our documentation model points the same way on a steeper gradient, although its outcome is documented-case
status rather than a documentation rate among known-positive drivers, so the two are not the
same quantity. Its odds ratio is \orInjured{} for whether the driver herself was
recorded as injured, and it does so for an attribute that no field on the form requests. The
methodological literature anticipates a shortfall of this shape. \citet{gibbons2014measuring}
distinguishes
under-ascertainment from under-reporting, and what we measure is under-reporting in the strict
sense, since the person reached the system and the attribute did not. \citet{german2001updated}
names sensitivity as a property of a surveillance system, and the \citet{international1995capture}
formulation makes clear why a single-source estimate needs an external expectation in place of
list overlap.

Set against the clinical literature on model-extracted variables, the extractor performs as that
literature predicts for a primary presence question. A weighted sensitivity of \seVal{} and
specificity of \spVal{} on a blind adjudicated sample sits at the upper end of the range reported
for binary clinical tasks. Zero-shot grading of postoperative complications reached 0.93 to 0.94
against a two-clinician benchmark \citep{warmer2026zeroshot}, and medication-error
classification reached binary concordance of 96.0\% \citep{krifors2026improving}. The consistent
finding in that literature is that performance degrades on finer attributes, falling to 76.5\% on
subcategorization in that same medication study \citep{krifors2026improving} and to moderate
agreement on secondary attributes in adverse-event extraction \citep{guillot2026extracting}. Our design accepts that finding rather
than testing against it, which is why the count rests on the presence question alone and why
stage and role are reported as descriptive attributes with no correction applied to them. Against
the crash-narrative literature the comparison is one of purpose rather than accuracy.
\citet{su2026llmdriven} reached 86.6\% accuracy on causal-frame extraction from California
narratives, \citet{zhang2026} compared eleven models on 16{,}656 Kentucky narratives, and
\citet{crashsage2025} outperformed tabular baselines on severity. Each returns a label, and a
label carries no error rate, so none of them can be assembled into a population count without an
assumption this paper measures instead. \citet{arteaga2025a} is the closest precedent in
intention, since it also used narratives to expose under-reporting in coded fields, and the step
we add is the correction and the denominator. The companion methods paper audits the same
instrument across twenty-seven variables and supplies the calibration argument that this paper
uses only for one.

The documented fetal-harm series should be read against the clinical case literature with the
difference in kind held firmly in view. We record \nFetalHarm{} crashes in which a narrative
states harm to or loss of the unborn child, spread across nine years and four stage groups at
single-digit counts throughout. The clinical sources count something rather different from that. \citet{weiss2001}
counted 240 certified traumatic fetal deaths across sixteen states over three years, of
which crashes caused 82\%. The meta-analysis of \citet{amezcuapriet2020maternal} reports
pooled incidences per 1{,}000 crash-involved women of 276.43 for induced labor, 191.90 for
preterm delivery and 166.65 for cesarean section. A scene-documented
count and a clinical outcome differ in three ways that all point the same direction. The officer
writes at the scene or shortly after, so a loss diagnosed days later never enters the narrative.
The officer records what was said rather than what was found, and the case reported by
\citet{matsushita2014fetal} shows fetal intracranial injury emerging weeks after a crash in which
the mother had no major injury. The narrative is also written for a legal purpose rather than a
clinical one. Our series is therefore a lower bound on documented harm and a far lower bound on
occurring harm, and its value lies in showing that such events reach the police record at all and
in what circumstances, rather than in estimating how often they happen.

The practical implication is a change to the form rather than to the analysis. Roughly a third of
the documented cases are passengers, whom the coded person file in this extract cannot even
count, and the documentation model shows that the cases which do get recorded are the ones in
which somebody was already hurt. A single pregnancy checkbox on the crash report, or a structured
field for occupant condition, would convert a signal that is laborious to recover into an ordinary coded variable, and the estimates here give the size of what such a field would capture. The
clinical guidance already treats pregnancy as a field-level fact worth establishing, because
management protocols turn on gestational age and on the need to monitor a patient whose injuries
look minor \citep{barraco2010practice,jain2015guidelines,greco2019guidance}. Trauma affects
between six and seven percent of pregnancies, with as much as 46\% of maternal deaths
attributable to it \citep{chames2008trauma}. An emergency medical technician who asks the question at the
scene generates exactly the information the narrative currently carries by accident. The same
argument reaches prenatal care from the other side. Counseling on restraint use ranged from
36.7\% to 56.5\% across nineteen states \citep{beck2005prevalence}, fewer than half of pregnant
women report receiving it \citep{sirin2007seat,taylor2005seatbelt}, and among 1{,}491 Australian
pregnant drivers 99.1\% wore a belt while only 41.4\% positioned it correctly
\citep{koppel2025assessing}. Our own comparison is consistent with that picture and adds one
reassuring detail, since unbelted status among documented pregnant drivers is 0.9\% against 0.9\% among
the sole female drivers aged 15 to 49 who make up the model's control set. Correct positioning is what the counseling literature says is
missing, and a crash report cannot observe it, which marks the boundary of what this method can
contribute.

Replication elsewhere is the contribution we would most like to see taken up, and what it
demands is why it is feasible. The schema is eight questions, the screen is a regular
expression, the model call is a single pass over each narrative, and the validation sample is a
few hundred narratives read blind. The workflow is proposed for any state whose crash file pairs a free-text narrative with a
person-level record joinable by crash identifier. It has been run in one state, against one
reporting environment and one pinned model version, so transfer and effort are claims a
second state would have to demonstrate. The
documentation model needs only coded injury status and a narrative, so it can be fitted wherever
those two fields coexist. A multi-state comparison of documentation practice would be more informative
than any single state's count, because the quantity that varies across states is the one this
paper measures. The county pattern already suggests how much room there is for that variation,
since the smoothed documentation rate among the ten largest Texas counties spans a factor of
\countyTopSpread{} and nothing in that spread tracks any plausible geography of pregnancy. Two
practical cautions apply to any replication of this work. The first is de-identification, because the rate at
which identifiers survive a supplier's redaction determines what a keyword screen can still see
and what a coder may be shown. The second is the person file, since an extract that holds one row
per unit rather than one per person will reproduce the passenger gap described in
Section~\ref{sec:data} and will limit rate estimation to drivers, as it did here.

\section{Conclusion}\label{sec:conclusion}

Reading a state's entire police narrative corpus with a calibrated decision model recovers a
variable that no crash database codes. Texas narratives document a pregnancy in \nAdjustedTotal{} police-reported
crashes over the nine years from 2017 through 2025, with a 95\% interval of
\nAdjustedLo{} to \nAdjustedHi{}, including \nFetalHarm{} crashes in which the narrative records
harm to or loss of the unborn child. Expressed against person-level denominators the count
becomes a rate of \rateMin{} to \rateMax{} per 1{,}000 female drivers of reproductive age
involved in crashes. That rate falls by about \trendPct{}\% a year across the period,
with an interval of \trendRRLo{} to \trendRRHi{} on the annual rate ratio.
Expressed against the involvement that Texas vital statistics lead one to expect, the same count
becomes a surveillance sensitivity of \survMin{}\% to \survMax{}\% under a live-birth
expectation. The schema, the code and the aggregated outputs are released so that the
same comparison can be made in any state whose crash file carries a narrative.

Four contributions follow from the results set out above. The extractor is validated rather than
asserted, with a
weighted sensitivity of \seVal{} and specificity of \spVal{} measured against blind human
adjudication of a probability-stratified sample, and with the measured error rates entering the
count through a correction rather than sitting beside it as reassurance. The counts are corrected
in two directions at once, for classifier error inside the screened stratum and for screen recall
outside it. Both corrections travel through one bootstrap chain, so the reported intervals are
intervals on the whole estimation procedure rather than on its last step. The
documentation model turns a count into a statement about the reporting system, and the identity
of Equation~(\ref{eq:identity}) says what a complete version of that turn would need, because
surveillance sensitivity is exactly the average documentation probability over the pregnant
population. The descriptive epidemiology then shows what the documented population looks like,
and its most useful feature is a negative one. Restraint use among documented pregnant drivers is
indistinguishable from restraint use among the sole female drivers aged 15 to 49 who make
up the model's control set. That comparison is between documented pregnant drivers and
that control set rather than between pregnant and non-pregnant drivers, and documented cases are selected on injury, which restraint use
itself affects.

\limitationsparagraph{}The work carries several limitations and they compound in one direction
rather than cancelling. Documentation is not prevalence, and every count reported here is a
documentation count whose dependence on the officer, the agency and the clinical relevance of
pregnancy at the scene is measured in Section~\ref{sec:res-doc} but cannot be removed by
measuring it. Fetal harm is documented at the scene or in immediate follow-up rather than
diagnosed, so the series of \nFetalHarm{} events is not a clinical outcome and almost certainly
undercounts loss identified later. Validation rests on a few hundred adjudicated narratives
rather than thousands, and the specificity of \spVal{} reflects zero false positives among
twenty-seven adjudicated true negatives rather than a demonstration that the extractor never
errs. That is why the bootstrap draws the quantity from a posterior instead of fixing it. The
adjudicated share behind the screen-miss term rests on \nStageCReviewed{} Stage-C positives and
the miss term scales linearly with it, so that sample is the first thing a replication should
enlarge. Whether the extractor itself misses pregnancies among the unflagged narratives was
tested by reading the bands nearest its threshold entire, and none was found. Below that the narratives all carry the model's smallest reported value, and every one of
them was searched for a wider pregnancy vocabulary rather than sampled; what bounds that
region is the conjunction of two vocabularies failing, not a measured rate. The person extract holds one row per unit, so no passenger denominator exists and no
passenger rate is reported even though roughly a third of documented cases are passengers. Rows whose age is missing are dropped from the denominators. Expected counts are built from live births alone, which omits pregnancies ending otherwise
and makes the reported share too large. They also assume crash involvement is independent
of pregnancy, and the literature reports a second-trimester increase in serious crashes
rather than a ratio above one across a whole pregnancy, so that second departure has no
established direction here. The expectation is standardized to the drivers' own age distribution using Texas
age-specific birth rates, which are published for \ageStdYears{} of the nine years; the
remaining years carry an interpolated factor, and 2024 and 2025 hold the last observed one. The corpus
retains residual identifiers after two redaction passes, at \pctPIIExcluded{}\% among screened
fetal-harm narratives, which constrains what can be displayed here and should constrain
replication elsewhere. The model is closed and pinned to the single version
\texttt{jev-1.13.0}, recorded on every call, so results are not guaranteed to reproduce against a
later version. The study covers one state across nine years, with 2025 only partially reported, and the
2025 births and population behind its expected count are carried forward and extrapolated
rather than observed.

What the work makes possible is more specific than more research. The adjudicated Stage-C
sample is the cheapest thing to enlarge and the most valuable, because it is the one term whose
uncertainty scales linearly into the total. A person file carrying passengers rather than units
would extend rate estimation to the third of documented cases this paper can only count, and that
is an administrative change rather than a scientific one. Linkage to hospital discharge or birth
records would convert the surveillance figure from a live-birth scenario into a measurement and
would answer the independence question that vital statistics alone cannot. And because the
documentation model needs nothing but coded injury status and a narrative, it can be fitted
wherever those two coexist, which is most of the country; whether it transfers is a question
several states could settle together.

The most direct test is also the simplest. A pregnancy field on a crash report, or a structured
field for occupant condition, would turn a signal that is laborious to recover into an ordinary
coded variable. Until one exists, the narrative is the only place the information lives. This paper shows
that it can be read at the scale of a state, corrected for the errors of reading it, and
reported with intervals that carry those errors rather than setting them aside.

\appendix
\renewcommand{\thetable}{\thesection.\arabic{table}}
\renewcommand{\thefigure}{\thesection.\arabic{figure}}
\renewcommand{\theequation}{\thesection.\arabic{equation}}
\renewcommand{\theHtable}{\thesection.\arabic{table}}
\renewcommand{\theHfigure}{\thesection.\arabic{figure}}
\renewcommand{\theHequation}{\thesection.\arabic{equation}}

\FloatBarrier
\section*{Appendices}

\section{Case ascertainment}\label{app:regex}
\setcounter{table}{0}\setcounter{figure}{0}\setcounter{equation}{0}

The Stage-A screen is a regular expression rather than a classifier, and it was built to be
generous on the understanding that a model would do the discrimination afterward. Its vocabulary
was assembled in three passes. The first pass took the pregnancy-specific words that any writer
describing a pregnant person would be likely to reach for, namely the stem of pregnant itself
together with unborn, fetus, fetal, trimester, gestation and the obstetric vocabulary of
placental injury and miscarriage. The second pass read a sample of the narratives those words
returned and added the phrasings officers actually use, which are shorter and more colloquial
than the clinical vocabulary. Those additions include a bare statement of weeks or months along
and a reference to labor and delivery as a destination rather than as an event. The third pass added a tier of
everyday English that carries an obstetric reading only sometimes, namely expecting, with child,
contractions and due date. That third tier is where the design decision sits. Each of those words
is far more common in its ordinary sense than in its obstetric one, and a screen built to
maximize precision would drop all four. Including them adds no work, because the screen is a regular
expression and the model reads each narrative once however many terms matched it. The tier
converts a precision problem that a regular expression cannot solve into a recall gain that
the model can then filter.

Table~\ref{tab:regex} gives every term, its tier, the pattern as it was applied, and three
counts. The first count is the narratives the term matched. The second is the narratives it was
the only matching term for, which measures what would be lost by dropping it. The third is the
narratives it matched that no core-tier term matched at all, which is the quantity that justifies
the everyday-English tier. Matching is case-insensitive and is done on substrings, so the pattern
\texttt{pregnan} catches pregnant, pregnancy and pregnancies without three separate entries. Each
term was applied separately as well as in combination, which is why the per-term counts are
recoverable at all; a single combined pass would have produced the hit set and nothing else. The
counts show a very uneven distribution of work. One term carries the screen almost single-handed, and
four terms return nothing that other terms do not already return, two of those four
returning nothing whatever. The everyday-English tier behaves as a nearly disjoint
addition, since almost every narrative it matches carries no core term. A replication in another state should expect the same shape and should rebuild the second
pass against its own narratives, because the colloquial register of police writing is local
in a way the clinical vocabulary is not.

The screen's recall was checked from the other side as well. Every narrative the extraction
model scored at its floor was searched for any of \nSecondOrderTerms{} further terms this
vocabulary does not carry, including maternity, prenatal, postpartum, in utero, gravid,
amniotic, midwife, stillborn, caesarean and sonogram; \nFloorFrame{} of \nFloorBand{}
matched and neither states a pregnancy. One term had to be dropped from that list first. A
bare \texttt{ob} matched several hundred narratives and every one was the direction
outbound, as in Southwest Freeway OB, which is why the screen's own pattern requires the
\texttt{gyn}
and is the kind of false friend a replication should expect to meet.

\begin{table}[htbp]
\rmfamily\centering
\small
\caption{Regular-expression families of the Stage-A screen, with per-term hit counts over the 5,018,079 narratives of the corpus.}
\label{tab:regex}
\begin{threeparttable}
\begin{tabular}{>{\raggedright\arraybackslash}p{2.4cm}>{\raggedright\arraybackslash}p{2.3cm}>{\raggedright\arraybackslash}p{3.2cm}rrr}
\toprule
\hdr{Term} & \hdr{Tier} & \hdr{Pattern} & \hdr{Narratives\nl{}matched} & \hdr{Matched by\nl{}this term alone} & \hdr{Matched with\nl{}no core term} \\
\midrule
\texttt{pregnan} & Core & \texttt{pregnan} & 5,318 & 5,112 & 0 \\
\texttt{unborn} & Core & \texttt{unborn} & 123 & 38 & 0 \\
\texttt{fetus} & Core & \texttt{fetus} & 31 & 8 & 0 \\
\texttt{fetal} & Core & \texttt{fetal} & 29 & 25 & 0 \\
\texttt{ob\_\allowbreak{}gyn} & Core & \texttt{ob[/\textbackslash{}-\textbackslash{}. ]?gyn} & 17 & 4 & 0 \\
\texttt{miscarr} & Core & \texttt{miscarr} & 15 & 8 & 0 \\
\texttt{trimester} & Core & \texttt{trimester} & 13 & 1 & 0 \\
\texttt{labor\_\allowbreak{}and\_\allowbreak{}delivery} & Core & \texttt{labor and delivery|labour and delivery} & 8 & 5 & 0 \\
\texttt{placent} & Core & \texttt{placent} & 6 & 3 & 0 \\
\texttt{gestation} & Core & \texttt{gestation} & 3 & 3 & 0 \\
\texttt{weeks\_\allowbreak{}along} & Core & \texttt{weeks along} & 2 & 0 & 0 \\
\texttt{months\_\allowbreak{}along} & Core & \texttt{months along} & 1 & 0 & 0 \\
\texttt{foetus} & Core & \texttt{foetus|foetal} & 0 & 0 & 0 \\
\texttt{obstetric} & Core & \texttt{obstetric} & 0 & 0 & 0 \\
\texttt{expecting} & Everyday English & \texttt{expecting} & 864 & 862 & 862 \\
\texttt{with\_\allowbreak{}child} & Everyday English & \texttt{with child} & 544 & 541 & 541 \\
\texttt{contractions} & Everyday English & \texttt{contractions} & 76 & 19 & 19 \\
\texttt{due\_\allowbreak{}date} & Everyday English & \texttt{due date} & 6 & 2 & 2 \\
\bottomrule
\end{tabular}
\begin{tablenotes}[flushleft]\footnotesize
\item[] Note: a term matches case-insensitively as a substring, so pregnan also catches pregnancy and pregnancies. The second count gives narratives this term alone matched, the third those no core term matched, which is what the everyday-English tier adds. The tiers together hit 6,840 narratives, 5,416 on a core term and 1,424 on everyday English alone.
\end{tablenotes}
\end{threeparttable}
\end{table}

\FloatBarrier
\section{The extraction schema}\label{app:schema}
\setcounter{table}{0}\setcounter{figure}{0}\setcounter{equation}{0}

The schema is the whole of the instrument, in the sense that it is the only thing separating this
analysis from a keyword count. It holds eight questions of three types. A \texttt{noul} question
states a proposition about the narrative and receives a probability that the proposition is true.
A \texttt{choice} question names a set of mutually exclusive options and receives a probability
distribution over them. A \texttt{score} question names ordered levels and receives a
distribution over those. No question invites free text, and the model returns no text, so every
answer is a typed value with a probability attached and nothing has to be parsed out of a
sentence after the fact.

Three of the eight questions are gated on the presence question, which is the design feature that
keeps the schema coherent. Asking what stage of pregnancy a narrative states, of a narrative that
states no pregnancy at all, has no defensible answer, and a model forced to choose among stages
will choose one. Gating forces those three questions to their \texttt{no\_pregnancy} option
whenever the presence probability is at or below the threshold, so the analysis never inherits a
role or a stage that rests on a pregnancy the model did not find. The ungated answers are
retained in separate columns, and the difference between the gated and ungated coding is one of
the sensitivity checks carried in the released outputs. The four remaining questions are asked of
every narrative regardless, because transport, a driver medical episode, narrative injury
severity and
a cited witness are all meaningful whether or not anyone was pregnant, and two of them serve as
consistency checks on the gated answers.

Table~\ref{tab:schema} reproduces all eight questions with their complete instruction and
criteria text, exactly as the model received it and exactly as the human coders received it.
Those two audiences read the same words on purpose. A disagreement between a coder and the model
is informative only if both were told the same thing, and a validation study in which the human
instructions paraphrase the model instructions measures the paraphrase as much as the model. The
criteria text is given in full rather than summarized, because the wording of a criterion is the
operational definition of the variable. A reader who wants to know what this paper counts as a
documented pregnancy will find the answer there rather than in the prose.

{\small
\begin{longtable}{>{\raggedright\arraybackslash}p{2.1cm}>{\raggedright\arraybackslash}p{1.1cm}>{\raggedright\arraybackslash}p{1.8cm}>{\raggedright\arraybackslash}p{3.9cm}>{\raggedright\arraybackslash}p{5.1cm}}
\caption{The \texttt{pregnancy\_v1} extraction schema in full, with every instruction and answer criterion as the model received it.}\label{tab:schema}\\
\toprule
\hdr{Question} & \hdr{Type} & \hdr{Gated on} & \hdr{Instruction as\nl{}given to the model} & \hdr{Answer criteria as\nl{}given to the model} \\
\midrule
\endfirsthead
\toprule
\hdr{Question} & \hdr{Type} & \hdr{Gated on} & \hdr{Instruction as\nl{}given to the model} & \hdr{Answer criteria as\nl{}given to the model} \\
\midrule
\endhead
\bottomrule
\endfoot
\bottomrule
\multicolumn{5}{@{}p{\linewidth}@{}}{\footnotesize Note: a gated question is forced to its no-pregnancy option whenever the presence probability is at or below the threshold; the ungated answer is kept for the sensitivity analysis. A \texttt{noul} question returns the probability that the stated proposition is true, a \texttt{choice} question a probability over the named options, and a \texttt{score} question a probability over ordered levels.}\\
\endlastfoot
\texttt{preg\_\allowbreak{}mentioned} & noul & not gated & The \textasciigrave{}narrative\textasciigrave{} states that a person involved in the crash was pregnant. & \textbf{true} Any explicit statement that a driver, passenger, pedestrian, or other involved person is pregnant, e.g. \textquotesingle{}was 6 months pregnant\textquotesingle{}, \textquotesingle{}pregnant driver\textquotesingle{}, \textquotesingle{}expecting\textquotesingle{}. \textbf{false} No pregnancy is stated. Mentions of children, infants, or car seats do not count. \\
\texttt{preg\_\allowbreak{}role} & choice & \texttt{preg\_\allowbreak{}mentioned} & What was the role of the pregnant person in the crash according to the \textasciigrave{}narrative\textasciigrave{}? & \textbf{driver} The pregnant person was driving one of the vehicles. \textbf{passenger} The pregnant person was a passenger in one of the vehicles. \textbf{pedestrian\_or\_other} The pregnant person was a pedestrian, cyclist, or otherwise not inside a vehicle. \textbf{no\_pregnancy} The narrative does not state that anyone was pregnant. \\
\texttt{preg\_\allowbreak{}outcome} & choice & \texttt{preg\_\allowbreak{}mentioned} & What does the \textasciigrave{}narrative\textasciigrave{} state about the condition of the pregnant person after the crash? & \textbf{no\_complaint} The pregnant person reported no injury or the narrative gives no complaint. \textbf{pain\_or\_evaluation} The pregnant person complained of pain, tightness, or discomfort, or was checked by EMS but not transported. \textbf{transported} The pregnant person was transported to a hospital or medical facility as a precaution or for injury. \textbf{fetal\_harm} The narrative states harm to, or loss of, the unborn child. \textbf{no\_pregnancy} The narrative does not state that anyone was pregnant. \\
\texttt{preg\_\allowbreak{}stage} & choice & \texttt{preg\_\allowbreak{}mentioned} & What stage of pregnancy does the \textasciigrave{}narrative\textasciigrave{} state for the pregnant person? & \textbf{early} Up to 13 weeks, or 1 to 3 months pregnant, or described as first trimester or early pregnancy. \textbf{mid} 14 to 27 weeks, or 4 to 6 months pregnant, or described as second trimester. \textbf{late} 28 weeks or more, or 7 to 9 months pregnant, or described as third trimester, due soon, or full term. \textbf{stage\_not\_stated} Pregnancy is stated but no weeks, months, or trimester is given. \textbf{no\_pregnancy} The narrative does not state that anyone was pregnant. \\
\texttt{transported} & noul & not gated & The \textasciigrave{}narrative\textasciigrave{} states that a person was transported to a hospital or medical facility. & \textbf{true} A person was transported by EMS, ambulance, helicopter, or private vehicle to a hospital, medical center, or clinic. \textbf{false} No transport to a medical facility is mentioned, or people refused transport. \\
\texttt{medical\_\allowbreak{}episode} & noul & not gated & The \textasciigrave{}narrative\textasciigrave{} states that a driver experienced a medical event (seizure, fainting, blackout, diabetic episode, heart attack, stroke, or similar) before or during the crash. & \textbf{true} A medical event of the driver is described as occurring before or during the crash, whether confirmed or suspected by the officer. \textbf{false} No medical event of the driver is described. Injuries caused by the crash, or transport to a hospital for crash injuries, do not count. \\
\texttt{injury\_\allowbreak{}narr} & score & not gated & What is the most severe injury outcome stated in the \textasciigrave{}narrative\textasciigrave{} for any person? & \textbf{0} No injury is mentioned, or the narrative says no one was injured. \textbf{1} A person complained of pain or possible injury but was not transported. \textbf{2} A person had a visible injury or was transported to a hospital. \textbf{3} A person was pronounced dead or the narrative describes a fatality. \\
\texttt{witness\_\allowbreak{}cited} & noul & not gated & The \textasciigrave{}narrative\textasciigrave{} cites a statement from a witness who was not a driver or occupant of the involved vehicles. & \textbf{true} A witness, bystander, or independent observer\textquotesingle{}s statement is described. \textbf{false} Only driver, passenger, or officer statements appear, or no statements appear. \\
\end{longtable}
}

\section{Validation protocol}\label{app:validation}
\setcounter{table}{0}\setcounter{figure}{0}\setcounter{equation}{0}

The validation sample exists to estimate two error rates, and its design follows from the fact
that those two rates are informative in different places. Sensitivity is determined by narratives
the model calls negative that adjudication calls positive, and specificity by narratives it calls
positive that adjudication calls negative. Both kinds live overwhelmingly in the middle of the
probability distribution, and the middle of the distribution is nearly empty, because the model
is decisive about most narratives. A simple random sample of the hit stratum would therefore have
spent most of its reading effort confirming the obvious. The sample is instead stratified on the
model's own probability, with the band from 0.30 to 0.95 over-sampled heavily, a hundred draws
taken from the region above 0.95 and twenty-five from the region below 0.30. The high tail
receives more draws than the low tail for a reason worth stating. Fifteen draws there would leave a single mislabeled item at a design weight above three
hundred carrying most of the weighted false-positive mass, which would make the
specificity interval a report on one reader's attention rather than on the extractor. Every Stage-C candidate is adjudicated as a census of its
own population, so those strata carry a weight of one.

Table~\ref{tab:weights} gives the resulting design and the weights it implies. Each rate in the
paper is a Horvitz-Thompson
ratio of weighted cell counts, computed by Equation~(\ref{eq:ht}), so the over-sampling raises
precision in the region that determines the answer without moving the estimate away from the
population it describes. The unweighted rates are reported beside the weighted ones in
Table~\ref{tab:confusion} so that a reader can see how much work the weights are doing, and the
bootstrap resamples within stratum on every draw so that the design survives into the interval.

Coders worked in a browser interface reached by a single-use access code, which displayed one
narrative at a time, held no crash identifier, offered no download or copy affordance, and was
revoked when labeling finished. Blind mode was on by default and hid the model's answer while the
narrative was read, and a coder could reveal it only after committing a label. Presentation order
was shuffled. The instructions given to coders are the criteria text of
Table~\ref{tab:schema}, reproduced word for word, plus a single governing rule, which is to code
only what the officer actually wrote and to record uncertainty rather than infer. Uncertainty was
offered as a first-class answer rather than treated as a failure, and rows marked uncertain are
set aside rather than scored, which is why the pair count behind the agreement statistic is
smaller than the number of double-read narratives. Coders were told that roughly a third of the
narratives they would read still carried a name, an address or another identifier despite two
redaction passes. They were asked to keep the file on their own machine and to delete it once
their labels had been returned, and they were covered by the data agreement for the duration.

Table~\ref{tab:confusion} gives the confusion cells in both forms together with every rate
derived from them. The specificity of one deserves the reading that the cells rather than the
rate support. It rests on twenty-seven adjudicated true negatives with no false positive among
them, which bounds the false-positive rate loosely rather than establishing that the extractor
cannot err. That is the reason the bootstrap draws specificity from the Jeffreys posterior of
Equation~(\ref{eq:beta}) rather than treating the point estimate as known.

\begin{table}[htbp]
\rmfamily\centering
\small
\caption{Sampling design of the validation sample, with the inverse-probability weight carrying each stratum back to its population.}
\label{tab:weights}
\begin{threeparttable}
\begin{tabular}{>{\raggedright\arraybackslash}p{6.6cm}rrrr}
\toprule
\hdr{Stratum} & \hdr{Narratives\nl{}in stratum} & \hdr{Narratives\nl{}drawn} & \hdr{Inclusion\nl{}probability} & \hdr{Design\nl{}weight} \\
\midrule
Stage-A hits, low tail $p \le 0.30$ & 1,467 & 25 & 0.0170 & 58.7 \\
Stage-A hits, ambiguous band $0.30 < p \le 0.95$ & 292 & 60 & 0.2055 & 4.9 \\
Stage-A hits, high tail $p > 0.95$ & 5,081 & 100 & 0.0197 & 50.8 \\
Stage-C non-hits, model positive $p > 0.50$ & census & 45 & 1.000 & 1.0 \\
Stage-C non-hits, borderline $0.20 < p \le 0.50$ & census & 177 & 1.000 & 1.0 \\
\bottomrule
\end{tabular}
\begin{tablenotes}[flushleft]\footnotesize
\item[] Note: the ambiguous band is over-sampled by design, because that is where an error rate is most informative. Every rate is weighted by the reciprocal of the inclusion probability, so the over-sampling raises precision without moving the estimate. Stage-C candidates are a census of their own population, so their weight is one.
\end{tablenotes}
\end{threeparttable}
\end{table}

\begin{table}[htbp]
\rmfamily\centering
\small
\caption{Confusion cells and the rates computed from them, at $\tau=0.5$.}
\label{tab:confusion}
\begin{threeparttable}
\begin{tabular}{>{\raggedright\arraybackslash}p{7.0cm}rr}
\toprule
\hdr{Cell or rate} & \hdr{Adjudicated\nl{}narratives} & \hdr{Weighted to the\nl{}hit population} \\
\midrule
Model positive, adjudicated pregnant (true positive) & 146 & 5,304.9 \\
Model negative, adjudicated pregnant (false negative) & 1 & 4.9 \\
Model positive, adjudicated not pregnant (false positive) & 0 & 0.0 \\
Model negative, adjudicated not pregnant (true negative) & 27 & 1,476.7 \\
Sensitivity & 0.9932 & 0.99908 \\
Specificity & 1.0000 & 1.00000 \\
Positive predictive value & 1.0000 & 1.00000 \\
Negative predictive value & -- & 0.99672 \\
Adjudicated prevalence within the hit stratum, weighted & -- & 0.78240 \\
\bottomrule
\end{tabular}
\begin{tablenotes}[flushleft]\footnotesize
\item[] Note: the left column counts adjudicated narratives as read. The right applies the design weights of Table~\ref{tab:weights} to estimate what the same decisions give over the whole Stage-A hit stratum, and every rate in the paper comes from it. A specificity of one records zero false positives among the adjudicated true negatives, which is why the bootstrap draws specificity from the posterior those cells imply.
\end{tablenotes}
\end{threeparttable}
\end{table}

\section{The bias analysis}\label{app:bias}
\setcounter{table}{0}\setcounter{figure}{0}\setcounter{equation}{0}

The quantitative bias analysis exists to answer a single question. If pregnancy is documented at
rate
$\delta_1$ in serious crashes and rate $\delta_0$ in non-serious ones, and if the undocumented
pregnancies are therefore sitting in the control group, what association between pregnancy and
crash severity would the data have shown had every pregnancy been recorded. The analysis is a
reconstruction rather than a correction. It does not claim to know $\delta_1$ or $\delta_0$, and
it produces an envelope rather than a point estimate, so that a reader can see the full set of
conclusions the data admit and can decide which corner of the grid they find credible.

The mechanics matter, and one way of getting them wrong is instructive. Up-weighting the
observed cases, as though the undocumented pregnancies were thinly represented versions of
the documented ones, returns an envelope so narrow around the naive estimate that it
amounts to the claim that detection bias barely matters. That is
backward. The undocumented pregnancies are not thinly represented cases, they are records
currently counted as controls, and they are concentrated in exactly the low-severity strata where
no one had a reason to write pregnancy down. The reclassification therefore has to move weight
between groups rather than scale it within one. In severity stratum $s$ with $d_s$ documented
cases, a documentation probability of $\delta_s$ implies $d_s/\delta_s$ true pregnancies and
therefore $d_s(1/\delta_s - 1)$ misclassified cases to be moved out of the controls of that
stratum, after which the association is refitted by Equation~(\ref{eq:grid}).

\begin{table}[htbp]
\rmfamily\centering
\small
\setlength{\tabcolsep}{3.4pt}
\caption{Complete scenario grid of the quantitative bias analysis. Each row assumes a documentation probability in serious and in non-serious crashes, reclassifies the undocumented pregnancies out of the control group accordingly, and refits the association.}
\label{tab:biasgrid}
\begin{threeparttable}
\begin{tabular}{rrrr@{\hspace{9pt}}rrrr@{\hspace{9pt}}rrrr}
\toprule
\hdr{$\delta_1$, serious} & \hdr{$\delta_0$, non-serious} & \hdr{Adjusted\nl{}odds ratio} & \hdr{Implied true\nl{}pregnancies} & \hdr{$\delta_1$, serious} & \hdr{$\delta_0$, non-serious} & \hdr{Adjusted\nl{}odds ratio} & \hdr{Implied true\nl{}pregnancies} \\
\midrule
0.02 & 0.02 & 1.658 & 122,550 & 0.30 & 0.50 & 2.684 & 5,018 \\
0.02 & 0.05 & 4.310 & 51,630 & 0.30 & 0.70 & 3.764 & 3,667 \\
0.02 & 0.10 & 8.719 & 27,990 & 0.30 & 0.90 & 4.844 & 2,917 \\
0.02 & 0.20 & 17.517 & 16,170 & 0.30 & 1.00 & 5.384 & 2,654 \\
0.02 & 0.30 & 26.311 & 12,230 & 0.50 & 0.02 & 0.060 & 118,374 \\
0.02 & 0.50 & 43.921 & 9,078 & 0.50 & 0.05 & 0.156 & 47,454 \\
0.02 & 0.70 & 61.572 & 7,727 & 0.50 & 0.10 & 0.316 & 23,814 \\
0.02 & 0.90 & 79.262 & 6,977 & 0.50 & 0.20 & 0.638 & 11,994 \\
0.02 & 1.00 & 88.119 & 6,714 & 0.50 & 0.30 & 0.961 & 8,054 \\
0.05 & 0.02 & 0.622 & 119,940 & 0.50 & 0.50 & 1.609 & 4,902 \\
0.05 & 0.05 & 1.619 & 49,020 & 0.50 & 0.70 & 2.259 & 3,551 \\
0.05 & 0.10 & 3.279 & 25,380 & 0.50 & 0.90 & 2.909 & 2,801 \\
0.05 & 0.20 & 6.595 & 13,560 & 0.50 & 1.00 & 3.235 & 2,538 \\
0.05 & 0.30 & 9.904 & 9,620 & 0.70 & 0.02 & 0.043 & 118,324 \\
0.05 & 0.50 & 16.512 & 6,468 & 0.70 & 0.05 & 0.111 & 47,404 \\
0.05 & 0.70 & 23.117 & 5,117 & 0.70 & 0.10 & 0.226 & 23,764 \\
0.05 & 0.90 & 29.723 & 4,367 & 0.70 & 0.20 & 0.455 & 11,944 \\
0.05 & 1.00 & 33.028 & 4,104 & 0.70 & 0.30 & 0.686 & 8,004 \\
0.10 & 0.02 & 0.304 & 119,070 & 0.70 & 0.50 & 1.149 & 4,852 \\
0.10 & 0.05 & 0.793 & 48,150 & 0.70 & 0.70 & 1.614 & 3,501 \\
0.10 & 0.10 & 1.607 & 24,510 & 0.70 & 0.90 & 2.080 & 2,751 \\
0.10 & 0.20 & 3.237 & 12,690 & 0.70 & 1.00 & 2.314 & 2,488 \\
0.10 & 0.30 & 4.866 & 8,750 & 0.90 & 0.02 & 0.033 & 118,297 \\
0.10 & 0.50 & 8.118 & 5,598 & 0.90 & 0.05 & 0.086 & 47,377 \\
0.10 & 0.70 & 11.364 & 4,247 & 0.90 & 0.10 & 0.175 & 23,737 \\
0.10 & 0.90 & 14.607 & 3,497 & 0.90 & 0.20 & 0.354 & 11,917 \\
0.10 & 1.00 & 16.227 & 3,234 & 0.90 & 0.30 & 0.533 & 7,977 \\
0.20 & 0.02 & 0.151 & 118,635 & 0.90 & 0.50 & 0.894 & 4,825 \\
0.20 & 0.05 & 0.392 & 47,715 & 0.90 & 0.70 & 1.256 & 3,474 \\
0.20 & 0.10 & 0.796 & 24,075 & 0.90 & 0.90 & 1.619 & 2,723 \\
0.20 & 0.20 & 1.604 & 12,255 & 0.90 & 1.00 & 1.801 & 2,461 \\
0.20 & 0.30 & 2.414 & 8,315 & 1.00 & 0.02 & 0.030 & 118,287 \\
0.20 & 0.50 & 4.033 & 5,163 & 1.00 & 0.05 & 0.078 & 47,367 \\
0.20 & 0.70 & 5.651 & 3,812 & 1.00 & 0.10 & 0.158 & 23,727 \\
0.20 & 0.90 & 7.268 & 3,062 & 1.00 & 0.20 & 0.319 & 11,907 \\
0.20 & 1.00 & 8.075 & 2,799 & 1.00 & 0.30 & 0.480 & 7,967 \\
0.30 & 0.02 & 0.100 & 118,490 & 1.00 & 0.50 & 0.804 & 4,815 \\
0.30 & 0.05 & 0.261 & 47,570 & 1.00 & 0.70 & 1.131 & 3,464 \\
0.30 & 0.10 & 0.529 & 23,930 & 1.00 & 0.90 & 1.458 & 2,714 \\
0.30 & 0.20 & 1.066 & 12,110 & 1.00 & 1.00 & 1.622 & 2,451 \\
0.30 & 0.30 & 1.605 & 8,170 &  &  &  &  \\
\bottomrule
\end{tabular}
\begin{tablenotes}[flushleft]\footnotesize
\item[] Note: the naive estimate on the same binary outcome is 1.622, and the grid returns it along the diagonal where the two probabilities are equal. Both orderings are evaluated, including the cells in which non-serious crashes are better documented, because the documentation model estimates a joint event and does not identify the conditional documentation probability that would rule one direction out. The implied count is the number of pregnant drivers a scenario requires in the eligible population over the nine years, and a scenario needing more than the nine-year expectation of Table~\ref{tab:annual} is marked in the released grid as inadmissible.
\end{tablenotes}
\end{threeparttable}
\end{table}

\begin{figure}[tbp]\centering
\includegraphics[width=0.86\textwidth]{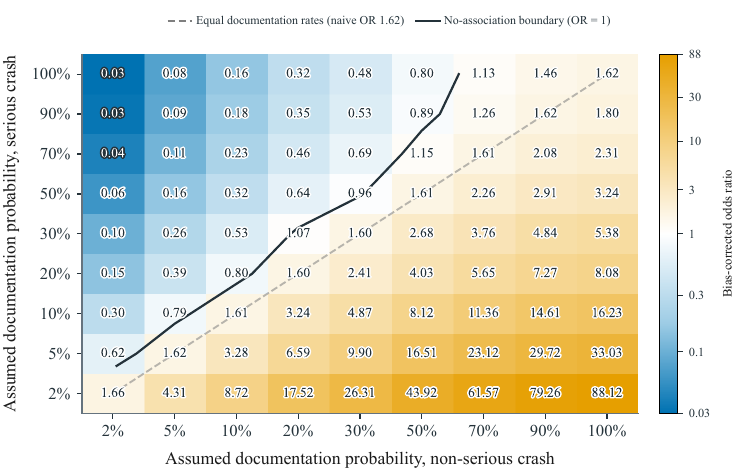}
\caption{The association between documented pregnancy and crash severity under detection bias.
Each cell assumes a documentation probability in serious crashes, read from the vertical axis,
and in non-serious crashes, read from the horizontal axis, reclassifies the undocumented
pregnancies out of the control group accordingly, and reports the odds ratio that results. The
dashed diagonal marks equal documentation in both strata, which returns the naive estimate, and
the solid contour marks the no-association boundary. Both orderings of the two strata are evaluated.}
\label{fig:bias}\end{figure}

Table~\ref{tab:biasgrid} gives every scenario in the grid and Figure~\ref{fig:bias} plots it.
\biasDocMin{} is the lowest documentation probability considered, chosen so that the grid
reaches the order of the surveillance estimate rather than stopping an order of magnitude
above it. The same nine values are used on both axes. All \biasGridN{} combinations
are evaluated, including those in which non-serious crashes are better documented than
serious ones. The documentation model of Table~\ref{tab:docmodel} does not rule that direction out,
because it estimates the joint event of being pregnant and being recorded rather than the
conditional documentation probability the grid assumes. The grid also reports the number of
pregnant drivers each scenario implies must exist in the eligible population, which is the
quantity that lets a reader reject a corner of the grid on external grounds. A scenario implying
more pregnant drivers than the vital-statistics expectation of Table~\ref{tab:annual} allows is
not admissible, and the implied counts are given so that comparison can be made rather than
asserted.

\FloatBarrier
\section{Derivations and the implementation map}\label{app:derivations}
\setcounter{table}{0}\setcounter{figure}{0}\setcounter{equation}{0}

This appendix carries the four short derivations that the methods section states as results,
together with the map from each numbered equation to the code that computes it. Nothing here
introduces a new estimate, and no quantity in this appendix is used anywhere in the paper except
as a check on a quantity computed elsewhere.

\begin{table}[htbp]
\rmfamily\centering
\small
\caption{Every numbered equation, the function implementing it, its script, and the file the result is written to.}
\label{tab:eqmap}
\begin{threeparttable}
\begin{tabular}{l>{\raggedright\arraybackslash}p{4.6cm}>{\raggedright\arraybackslash}p{2.6cm}>{\raggedright\arraybackslash}p{2.6cm}>{\raggedright\arraybackslash}p{3.0cm}}
\toprule
\hdr{Equation} & \hdr{Quantity} & \hdr{Function} & \hdr{Script or\nl{}appendix} & \hdr{Output file} \\
\midrule
(\ref{eq:ht}) & Horvitz-Thompson estimation of sensitivity and specificity & \texttt{weighted\_\allowbreak{}rates} & \texttt{p09\_\allowbreak{}validation.py} & \texttt{validation\_\allowbreak{}metrics.json} \\
(\ref{eq:rg}) & Corrected prevalence within the hit stratum & \texttt{rogan\_\allowbreak{}gladen} & \texttt{p10\_\allowbreak{}estimates.py} & \texttt{annual\_\allowbreak{}estimates.csv} \\
(\ref{eq:total}) & Two-stage adjusted count for one year & \texttt{main} & \texttt{p10\_\allowbreak{}estimates.py} & \texttt{annual\_\allowbreak{}estimates.csv} \\
(\ref{eq:beta}) & Jeffreys posterior for specificity at the zero-false-positive boundary & \texttt{beta\_\allowbreak{}from\_\allowbreak{}ci} & \texttt{p10\_\allowbreak{}estimates.py} & \texttt{estimates.json} \\
(\ref{eq:mc}) & Monte Carlo propagation through the estimator & \texttt{main} & \texttt{p10\_\allowbreak{}estimates.py} & \texttt{annual\_\allowbreak{}estimates.csv} \\
(\ref{eq:rate}) & Documented rate per thousand female drivers & \texttt{main} & \texttt{p10\_\allowbreak{}estimates.py} & \texttt{annual\_\allowbreak{}estimates.csv} \\
(\ref{eq:prev}) & Pregnancy point prevalence from the general fertility rate & \texttt{main} & \texttt{p07\_\allowbreak{}external.py} & \texttt{external\_\allowbreak{}vitals.json} \\
(\ref{eq:surv}) & Expected pregnant drivers and surveillance sensitivity & \texttt{main} & \texttt{p10\_\allowbreak{}estimates.py} & \texttt{annual\_\allowbreak{}estimates.csv} \\
(\ref{eq:doc}) & Documentation model as a logistic regression & \texttt{m1\_\allowbreak{}documentation} & \texttt{p12\_\allowbreak{}models.py} & \texttt{model\_\allowbreak{}documentation.csv} \\
(\ref{eq:identity}) & Surveillance sensitivity as a mean documentation probability & derivation & Appendix~\ref{app:derivations} & -- \\
(\ref{eq:trend}) & Negative binomial trend with an exposure offset & \texttt{m2\_\allowbreak{}trend} & \texttt{p12\_\allowbreak{}models.py} & \texttt{model\_\allowbreak{}trend.csv} \\
(\ref{eq:qba}) & Observed odds ratio under outcome-dependent documentation & derivation & Appendix~\ref{app:derivations} & -- \\
(\ref{eq:grid}) & Bias grid as the inverse map from documentation rates & \texttt{m3\_\allowbreak{}severity} & \texttt{p12\_\allowbreak{}models.py} & \texttt{model\_\allowbreak{}severity\_\allowbreak{}bias\_\allowbreak{}grid.csv} \\
(\ref{eq:eb}) & Gamma-Poisson empirical-Bayes posterior mean for a county & \texttt{m4\_\allowbreak{}counties} & \texttt{p12\_\allowbreak{}models.py} & \texttt{county\_\allowbreak{}rates.csv} \\
(\ref{eq:delta}) & Delta-method variance of the corrected prevalence & derivation & Appendix~\ref{app:derivations} & -- \\
(\ref{eq:orobs}) & Rare-exposure form of the observed odds ratio & derivation & Appendix~\ref{app:derivations} & -- \\
\bottomrule
\end{tabular}
\begin{tablenotes}[flushleft]\footnotesize
\item[] Note: scripts are under \texttt{paper2/src/} and outputs under \texttt{paper2/outputs/} or \texttt{paper2/data/} in the released repository. Four entries are derivations carried out in Appendix~\ref{app:derivations}; they check quantities the scripts produce rather than producing any.
\end{tablenotes}
\end{threeparttable}
\end{table}

\paragraph{The map from each numbered equation to the code that computes it.}
Table~\ref{tab:eqmap} names, for each numbered equation, the function that implements it, the
script that function lives in and the file its result is written to. The four rows without a script are the
derivations below. A reader checking this paper against its code should begin
there, because every number in the manuscript traces to one of those output files through the
macro file that the prose reads its numbers from.

\paragraph{A delta-method variance for the corrected prevalence within the hits.}
Write $J = \mathrm{Se} + \mathrm{Sp} - 1$ for the Youden index of the extractor and treat
$\hat q$, $\widehat{\mathrm{Se}}$ and $\widehat{\mathrm{Sp}}$ as independent, which the design
supports because the apparent prevalence is computed on the full hit stratum and the error rates
on a sample of it. Differentiating Equation~(\ref{eq:rg}) gives
$\partial\hat\pi/\partial\hat q = 1/J$, $\partial\hat\pi/\partial\widehat{\mathrm{Se}} =
-\hat\pi/J$ and $\partial\hat\pi/\partial\widehat{\mathrm{Sp}} = (1-\hat\pi)/J$, so to first
order
\begin{equation}\label{eq:delta}
\operatorname{Var}(\hat\pi) \;\approx\;
\frac{\operatorname{Var}(\hat q) + \hat\pi^{2}\operatorname{Var}(\widehat{\mathrm{Se}})
 + (1-\hat\pi)^{2}\operatorname{Var}(\widehat{\mathrm{Sp}})}{J^{2}} .
\end{equation}
Two things follow immediately. The variance is inflated by $1/J^{2}$, so an extractor with a
Youden index near one inflates the variance barely at all, and one with an index near zero
cannot be corrected at all. Within the numerator the sensitivity term is weighted by $\hat\pi^{2}$ and the specificity
term by $(1-\hat\pi)^{2}$. In the hit stratum the adjudicated prevalence is \prevInHits{},
and at the measured error rates the corrected $\hat\pi$ is barely distinguishable from it, so
the sensitivity term dominates by roughly an order of magnitude and the specificity term
contributes little. Over the whole corpus the weights reverse completely, $\hat\pi$ falls below one in a
thousand, and specificity becomes the only term that matters, which is the algebraic form of the
argument in Section~\ref{sec:val} for restricting the correction to the hit stratum.

\paragraph{Surveillance sensitivity written as a mean documentation probability over cases.}
Let $\mathcal{P}_t$ be the set of pregnant drivers involved in crashes in year $t$, with
$|\mathcal{P}_t| = E_t$ in expectation, and let $\delta(x) = \Pr(Y=1 \mid x)$ be the
documentation probability of Equation~(\ref{eq:doc}) at covariate value $x$. The documented count
is the sum of the documentation indicators over that set, so
$\mathbb{E}[\hat N_t\phi_t] = \sum_{i \in \mathcal{P}_t}\delta(x_i)
= E_t\,\mathbb{E}[\delta(x)]$, where the expectation on the right runs over the covariate
distribution of pregnant drivers rather than over the eligible population as a whole. Dividing by $E_t$ recovers Equation~(\ref{eq:identity}), namely $S_t =
\mathbb{E}[\delta(x)]$. The identity holds as algebra and is not estimated in this paper.
Its ingredient $\delta(x)$ is the probability that a pregnancy is written down given that
there is one. The model of Equation~(\ref{eq:doc}) is fitted on all eligible female drivers,
whose outcome is instead the joint event that a driver is pregnant and that the pregnancy
was recorded. Substituting the fitted probability for $\delta(x)$ gives a mean of a few
hundredths of a percent against a surveillance sensitivity of a few percent, which is the
arithmetic that shows the two are different quantities. Estimating
Equation~(\ref{eq:identity}) would need pregnancy status for drivers whose pregnancy was
never recorded.

\paragraph{The direction of the independence bias that enters the expected count.}
The expected count of Equation~(\ref{eq:surv}) multiplies female drivers by the population
prevalence of pregnancy, which assumes crash involvement and pregnancy are independent. Suppose
instead that pregnant women are involved in crashes at $\theta$ times the rate of women of the
same age who are not pregnant. The correct expected count is then $E_t^{\ast} = \theta E_t$ to
first order in the small prevalence $\rho_t$, and the correct surveillance sensitivity is
$S_t^{\ast} = S_t/\theta$. The live-birth construction of $\rho_t$ has a known direction, since pregnancies ending in
loss or termination are excluded from $B_t$ and their inclusion would raise $\rho_t$, raise
$E_t$ and lower $S_t$. The value of $\theta$ does not. \citet{redelmeier2014} report a rise
in serious crash risk during the second trimester against a pre-pregnancy baseline, which
is not the same as $\theta$ above one across a whole pregnancy and across all
police-reported crashes, and the fixed gestation of Equation~(\ref{eq:prev}) complicates
the comparison further. Only the first departure therefore signs the error, which is why
the surveillance figure is quoted throughout against a live-birth expectation rather than
as a bound.

\paragraph{The observed odds ratio when documentation itself depends on the outcome.}
In severity stratum $s$ let $N_s$ be the population total, $n_s$ the true pregnancies and $d_s =
\delta_s n_s$ the documented ones. The observed odds ratio compares documented pregnancies
against everyone else,
\begin{equation}\label{eq:orobs}
\mathrm{OR}_{\mathrm{obs}}
= \frac{\delta_1 n_1 / (N_1 - \delta_1 n_1)}{\delta_0 n_0 / (N_0 - \delta_0 n_0)}
= \frac{\delta_1}{\delta_0}\cdot\frac{n_1/(N_1 - \delta_1 n_1)}{n_0/(N_0 - \delta_0 n_0)} .
\end{equation}
When the exposure is rare in both strata, meaning $n_s \ll N_s$, the two denominators approach
$N_1$ and $N_0$ and the second factor approaches
$\mathrm{OR}_{\mathrm{true}}$, which gives Equation~(\ref{eq:qba}). The approximation is
comfortable here, because documented pregnancy accounts for about a tenth of a percent of the eligible
frame and the implied true counts in Table~\ref{tab:biasgrid} stay below a few percent of it in
every evaluated scenario. The practical content of the result is that the entire distortion is
carried by the ratio $\delta_1/\delta_0$ and not by the individual documentation rates. A reader
willing to state how much better serious crashes document pregnancy than non-serious ones has
thereby stated the correction, whatever they believe about the absolute level of documentation.

\FloatBarrier


\section*{Data and code availability}
The extraction schema \texttt{pregnancy\_v1}, the asynchronous runner and every analysis
script are released under the MIT license at
\url{https://github.com/pozapas/pregnancy-crash-narratives}. The figure and table generators, the blind adjudication tool and
the coder instructions are released with them.

The county table behind Figure~\ref{fig:geo} and the state rate it marks holds one row for each of the \nCounties{} Texas counties,
with its exposure, its raw rate, its smoothed rate and its shrinkage weight. That table is
supplied as supplementary material rather than as an appendix, because a table of that length
belongs beside the article rather than inside it. Model version \texttt{jev-1.13.0} is pinned
and is recorded on every call.

The crash narratives themselves are not released and cannot be. They are Texas Crash Records
Information System records held under an agreement that does not permit redistribution, and they
retain personal identifiers after de-identification at the rates Section~\ref{sec:data}
quantifies. The repository therefore documents how to obtain equivalent data rather than
supplying it. The public query tool and the bulk crash-data request form of the state department
of transportation do not return narrative text. The narrative forms part of the peace officer's
crash report, and Texas Transportation Code \S550.065(c) limits its release to parties connected
with the crash. A narrative extract at this scale therefore requires a data-sharing arrangement
with the department. The same pipeline runs against any state crash file that pairs a free-text
narrative with a person-level record joinable by crash identifier. The one property worth
checking first is how the supplier has de-identified the text, because that determines what a
keyword screen can still see.


\printcredits

\bibliographystyle{cas-model2-names}
\bibliography{references}

\end{document}